\documentclass[review, 3p]{elsarticle}

\usepackage{lineno,hyperref}
\usepackage{cite}
\usepackage{graphicx}
\usepackage{longtable, ltcaption}
\usepackage{multicol}
\usepackage{comment}
\usepackage{algorithm,algpseudocode}
\usepackage{mathtools}
\usepackage{subfigure}
\usepackage{color}
\usepackage{float}
\usepackage{multirow}
\usepackage{array}
\usepackage{amsmath}
\usepackage[justification=centering]{caption}
\usepackage{epsfig}
\usepackage{tikz}
\def\checkmark{\tikz\fill[scale=0.4](0,.35) -- (.25,0) -- (1,.7) -- (.25,.15) -- cycle;}
\usepackage{amssymb}
\usepackage{pifont}
\newcommand{\xmark}{\ding{55}}%

\modulolinenumbers[0]

\usepackage{numcompress}\biboptions{authoryear}

\begin{document}

\begin{frontmatter}

\title{Using shortest path to discover criminal community}



\author[mymainaddress,mysecondaryaddress]{Pritheega Magalingam\corref{mycorrespondingauthor}}
\ead{pritheega.magalingam@rmit.edu.au}

\author[mymainaddress]{Stephen Davis}

\author[mymainaddress]{Asha Rao}
\ead{asha@rmit.edu.au}

\cortext[mycorrespondingauthor]{Corresponding author. Tel.: +61 3 9925 1843}

\address[mymainaddress]{School of Mathematical and Geospatial Sciences, RMIT University,  Melbourne, Australia, GPO Box 2476, Melbourne, Victoria 3001.}
\address[mysecondaryaddress]{Advanced Informatics School, Level 5, Menara Razak, Universiti Teknologi Malaysia, Jalan Semarak, 54100 Kuala Lumpur, Malaysia.}

\begin{abstract}
Extracting communities using existing community detection algorithms yields dense sub-networks that are difficult to analyse. Extracting a smaller sample that embodies the relationships of a list of suspects is an important part of the beginning of an investigation. In this paper, we present the efficacy of our shortest paths network search algorithm (SPNSA) that begins with an `algorithm feed', a small subset of nodes of particular interest, and builds an investigative sub-network. The algorithm feed may consist of known criminals or suspects, or persons of influence. This sets our approach apart from existing community detection algorithms. We apply the SPNSA on the Enron Dataset of e-mail communications starting with those convicted of money laundering in relation to the collapse of Enron as the algorithm feed. The algorithm produces sparse and small sub-networks that could feasibly identify a list of persons and relationships to be further investigated. In contrast, we show that identifying sub-networks of interest using either community detection algorithms or a k-Neighbourhood approach produces sub-networks of much larger size and complexity. When the 18 top managers of Enron were used as the algorithm feed, the resulting sub-network identified 4 convicted criminals that were not managers and so not part of the algorithm feed. We also directly tested the SPNSA by removing one of the convicted criminals from the algorithm feed and re-running the algorithm; in 5 out of 9 cases the left out criminal occurred in the resulting sub-network.
\end{abstract}


\end{frontmatter}


\section{Introduction}
\label{Introduction}

Retrieving a criminal network from an organised crime incident is an important part of crime investigation. This task is a difficult one, mainly because of the involvement of a variety of criminals who play myriad roles \citep{basu2014, didimo11}. In addition to drug trafficking and money laundering, organised crime includes hijacking and equipment smuggling. The task of the criminal investigator is further hampered by the mass of data needing to be searched with an important part of the start of an investigation being the identification of a smaller sample that embodies the relationships within the criminal participants. In \citep{MagalingamP2014}, we presented an algorithm designed to extract a smaller, more manageable, network of possible relationships from a large dataset of interactions. In this paper, we further develop this algorithm and show that it performs well in a variety of scenarios, and is able to extract meaningful sub-networks for a criminal investigator to start an investigation. We show that this algorithm performs better than known community detection algorithms \citep{Pons20-06, Clauset20-04, Nman20-06}, as well as k-neighbourhood detection methods \citep{zhou2011}.

In the past, extracting criminal associations from raw data has required preliminary information of such relationships, while building a network from such, known, relationships has been done manually \citep{basu2014, didimo11, chris2010, oatley2014}. For example, \citet{nadji2013} produce a network of known fraudulent  infrastructure by creating links between IP addresses using known attack signatures garnered from passive domain name server and several other sources for malicious activities. \citet{krebs02} builds edges between known hijackers of the 9-11 terrorist attacks by manually gathering data from online news articles. The edges, or links, are created based of information such as whether the two persons went to the same school, grew up in the same locality, etc. \citet{oatley2014}  follow a similar track, using associations such as partner, sibling, cohabitant, to build  a relationship network among the members of different UK crime gangs. Clearly, the above methods are time-consuming, and a faster, more automated process of building a relationship network would be very useful for investigators of criminal activities.

We present such an algorithm, which can be run on a large dataset of interactions, to build a more practicable sub-network of known criminals suitable for further investigation. We use the publicly available Enron Dataset \citep{ISI:2009}, which contains all email communications before and after the collapse of this large company in 2001. This dataset is appropriate for this exercise, as ten people connected with Enron were subsequently convicted of money laundering \citep{secshamtransaction2004}.

The structure of the rest of the paper is as follows: In the next section, we describe the Enron dataset in more detail, give the process by which we start the isolation of specific email groupings, compare the connections between the ten criminals in two different email sub-networks,  and describe our algorithm. Section \ref{Discovery of Criminals} gives the results of applying existing community detection algorithms as well as the k-nearest neighbour method, to the Enron dataset to identify the community that the criminals belong. In the section after this, we apply our shortest paths network search algorithm to the two email sub-networks previously identified and compare the results to those obtained by applying the existing community detection algorithms. The penultimate section details the application of our algorithm to the different scenarios that an investigator may encounter.  Finally we give the conclusion.

\section{Background}
\label{background}

This section describes the preliminary analysis of the Enron email dataset, the people who were convicted of money laundering crime, the  identification of criminal communication links and the criminal sub-network formation methods.

\subsection{Preliminary analysis of dataset}
\label{Enron Email Dataset Description}

The Enron email dataset contains 1,887,305 email transactions \citep{ISI:2009} that were sent using the fields `TO', `CC' or `BCC'.  Out of these emails, 16,116 are senders of the emails and 68,203 are receivers of the emails. The Enron email dataset contains a mix of internal and external email transactions. Within the 16,116 email senders, 5,831 email transactions are from email addresses that are Enron company email accounts having the name \textit{`enron'} in their email address and the rest of the addresses are external, for example \textit{andrew.fastow@ljminvestments.com}, \textit{anitatr@earthlink.net}, etc. In order to process this large number of emails, we start by extracting the emails sent and received in the last 8 years of Enron - from 1995 to 2002 \citep{salter2008innovation}. We clean the data by removing the irrelevant email transactions such as email addresses that have numbers and characters for example `5673@aol.com', that end with airline company name for example `@aircanada.com', that end with `xpedia.com', `amazon.com' and other auto response emails.

Several prior works propose ways of extracting criminal networks in the form of associations between texts or people \citep{basu2014, krebs02}. Mining relevant terms from a large volume of police incident summaries and assigning the co-occurrence frequency as a weight to each term is used by \citep{che2004} to design a criminal network while \citet{yang2007} use web crawlers to gather identities associated with certain crime related topics in web blog pages and represent them as a network. Similarly, in order to identify criminal cliques, \citet{iqbal2012} perform chat topic analysis and certain entities that belong to the same chat session are formed into a clique. \citet{louis2011} conduct text mining on passages of a mystery novel to show the association between words, in the form a graph, leading to the identification of murders. In \citep{anwar2012}, posts that promote hate and violence in certain dark web forums are grouped in different cliques using an algorithm that measures similarity based on content, time, author and title.

Using keywords as a tool for isolating criminal networks is a problem especially when electronic documents, chat messages, web blogs or emails contain incomplete information or could mislead detection algorithms \citep{murynets2012, keila2005}. Consequently, we choose to ignore the content of the various emails being exchanged between the criminals and propose a very different way to start the building of a criminal network, by considering the type of emails based on recipient fields. As detailed in \citep{MagalingamP2014}, we separate the emails with at least one BCC recipient because the existence of a bcc in an email, could indicate a trust relationship \citep{fox2012}. While `to' and `cc' recipients are visible to all recipients, as \citep{uscertgovonBCC} and \citep{bogawar12} point out, there is something inherently secretive about adding a `bcc' recipient to an email. We explore the suspect's secret connections in the group of emails that consists of all those with one or more bcc-ed recipients and compare our result with the connections found using email transactions that have recipients only in the `TO' and `CC' fields.

\begin{figure}[H]
	\centering
	\subfigure[\small Degree distribution of `TO \& CC' email group ]
	{
	\includegraphics[height=1.4in, width=1.6in]{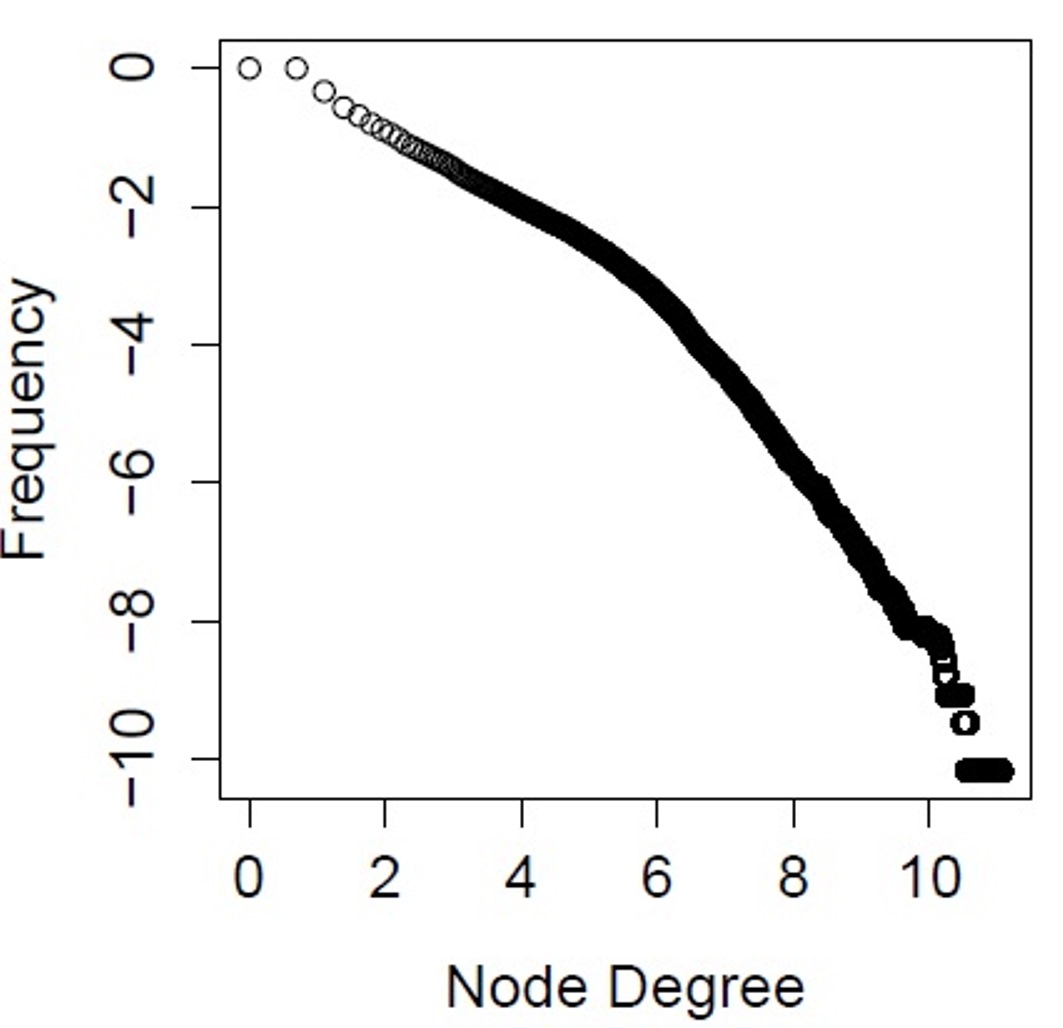}
	\label{fig:first_sub}
	}
	\hspace{1em}
	\subfigure[\small Degree distribution of `BCC' email group]
	{
		\includegraphics[height=1.4in, width=1.5in]{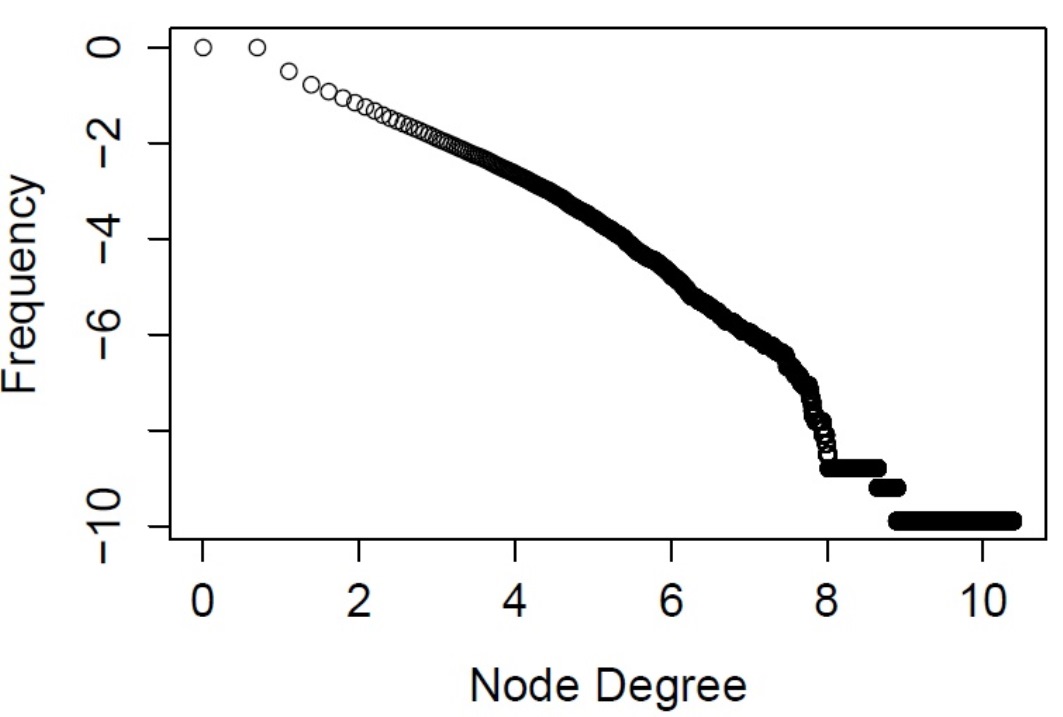}
		\label{fig:second_sub}
	}
      \captionsetup{singlelinecheck=off,justification=justified}
	\caption{\small The figures show that the degree distributions for both Netgraph (a) and BCC Netgraph (b) are heavy-tailed with many nodes being of low degree and some nodes being highly connected.}
	\label{fig:TOCCBCCEmailDegreeDist}
\end{figure}

The emails are divided into two groups. The first group is made up of email transactions that have recipients in the `TO' and `CC' fields. These emails do not contain any BCC recipients. The network formed using the `TO' and `CC' email transactions has 26,027 nodes and 1,048,572 edges with an average degree of 80.58. Henceforth, we refer to this network as Netgraph and each node ID is now called Net. ID. The relationship between the nodes' degrees and their frequencies is displayed in the log-log plot of the degree distribution (Figure \ref{fig:TOCCBCCEmailDegreeDist} (a)) .

\vspace*{-3mm}

The second group of emails consists of all those with one or more bcc-ed recipients. The network formed using the BCC email transactions is called the BCC Netgraph with each node in this BCC Netgraph given a BCCNet. ID. The BCC Netgraph contains 19,716 nodes and 238,761 edges. The BCC Netgraph has an average degree of 24.22 and the log-log plot of the degree distribution is shown in Figure \ref{fig:TOCCBCCEmailDegreeDist} (b). As is evident from Figure \ref{fig:TOCCBCCEmailDegreeDist} (a) and (b), both networks have many nodes with low degree and a few nodes with very high degree indicating the degree distributions are heavy tailed. Sub-networks are then constructed with each of these large networks using our shortest paths network search algorithm (see Section \ref{DescriptionofSPNSA}). Next the persons convicted of money laundering in relation to the collapse of Enron are listed along with their Net. IDs and BCCNet. IDs.

\vspace*{0 mm}

\subsection{Enron money laundering criminals}
\label{Money laundering criminals}

\vspace*{0 mm}

Ten people were convicted of money laundering in relation to the collapse of Enron  \citep{Kathleen2003}. Table \ref{table:MLCriminals} below shows the Net. ID of the criminals appearing within Netgraph as well as the BCCNet. ID of those appearing within the BCC Netgraph.

	\begin{table}[H]
		\protect\caption{{\small Enron money laundering criminals}}
			\label{table:MLCriminals}
		
		\vspace{0.2cm}
				
		\centering{}%
	
		\begin{tabular}{|l|>{\centering}m{1.2cm}|>{\centering}m{2.0cm}|>{\centering}m{5.5cm}|}
			\hline
			 \small Name & \small Net. ID &  \small BCCNet. ID &  \small Email Address\tabularnewline
			\hline
			\small Andrew Fastow & \small 1472  & \small 686 & \small andrew.fastow@enron.com\tabularnewline
			\hline
			\small Andrew Fastow & \small - & \small 687 & \small andrew.fastow@ljminvestments.com \tabularnewline
			\hline
			\small Lea Fastow & \small 17589 & \small 11010 & \small lfastow@pop.pdq.net \tabularnewline
			\hline
			\small Lea Fastow & \small 17588 & \small 11009 & \small lfastow@pdq.net\tabularnewline
			\hline
			\small Kevin Hannon & \small 16202 & \small 10068 & \small kevin.hannon@enron.com\tabularnewline
			\hline
			\small Kenneth Rice & \small 16115 & \small 9994 & \small kenneth.rice@enron.com\tabularnewline
	     	\hline
			\small Rex Shelby & \small 23983 & \small 15224 & \small rex.shelby@enron.com\tabularnewline
			\hline
			\small Rex Shelby & \small 23985 & \small 15225 & \small rex\_shelby@enron.net\tabularnewline
			\hline
			\small A. Khan & \small - & \small 205 &\small adnankkhan@hotmail.com\tabularnewline
			\hline
			\small Michael Kopper & \small 20217 & \small 12708 & \small michael.kopper@enron.com\tabularnewline
			\hline
			\small Ben Glisan & \small - & \small 1369 & \small ben.glisan@enron.com\tabularnewline
			\hline
			\small Joe Hirko & \small 14052 & \small 8716 & \small joe.hirko@enron.com\tabularnewline
			\hline
			\small S. Yaeger & \small - & \small 861 & \small anne.yaeger@enron.com\tabularnewline
			\hline	
		\end{tabular}
		\vspace{0.2cm}		
		\begin{footnotesize}
			\begin{minipage}{\linewidth}
				\small {Table \ref{table:MLCriminals} gives the list of e-mail accounts associated with criminals involved in the Enron money laundering crime \citep{Kathleen2003, ISI:2009}. The IDs in the table are computer generated numbers assigned to distinct email addresses based on the type of network. The Net. ID refers to the email addresses of criminals in the Netgraph while the BCCNet. ID refers to the email addressess of criminals in the BCC email network.}
			\end{minipage}
		\end{footnotesize}
		
	\end{table}

The multiple email addresses of the criminals, leading to multiple, different IDs, are preserved as some of these criminals, for example Andrew Fastow (BCCNet. ID 687), A. Khan (BCCNet. ID 205), Ben Glisan (BCCNet. ID 1369) and S. Yaeger (BCCNet. ID 861), did not occur in the Netgraph but were present in the BCC Netgraph. Each of these email addresses occur in distinct email transactions. We now identify the length of the shortest paths between these criminals in both the Netgraph and the BCC Netgraph.

\subsection{Distribution of criminal links in Netgraph and BCC Netgraph}
\label{IdentifyingNetworkMeasure}

Here, an analysis is conducted to compare the connections formed in the Netgraph and BCC Netgraph. Network measures are often used to quantify network structures, for example, the number of vertices in a network measures the size of the network, a vertex's degree could be used to show whether it is strong or weak, the shortest path length measures the distance between vertices, centrality measures demonstrate the level of importance of vertices, etc. \citep{newman2010}. We use the length of the shortest paths and the degree of the node. Both the Netgraph and the BCC Netgraph are directed graphs. In order to analyse a criminal's communication links within these graphs, we first calculate the length of the shortest paths from one criminal to another.  Tables \ref{table:SPLengthDNetgraph} and \ref{table:SPLengthDBCCNetgraph} shows the length of shortest paths from criminal to criminal in the directed Netgraph and the BCC Netgraph respectively. The directed Netgraph has an average path length 3.203258 while the directed BCC Netgraph's average path length is 4.445533.

\vspace*{10mm}

\begin{table}[H]
\centering
	\caption{Shortest path length from criminal to criminal in the directed Netgraph}
	\label{table:SPLengthDNetgraph} 
\begin{tabular}{l*{8}{l}l}
              \multicolumn{10}{c} {\footnotesize  Directed Netgraph} \\
	 {\tiny }  & {\tiny 1472} & {\tiny17589} & {\tiny 17588} & {\tiny 16202} & {\tiny 16115}  & {\tiny 23983} & {\tiny 23985} & {\tiny 20217} & {\tiny 14052} \\
	\hline
	{\tiny 1472}          & {\tiny 0} & {\tiny 3} & {\tiny 3} & {\tiny 4} & {\tiny 3} & {\tiny 2} & {\tiny 3 }& {\tiny 3 }& {\tiny 3}  \\
	{\tiny 17589}       & \tiny {Inf }& \tiny {0} & \tiny {Inf} & \tiny {Inf} &  \tiny {Inf} & \tiny {Inf} & \tiny {Inf} & \tiny {Inf} & \tiny {Inf}  \\
	{\tiny 17588}      \tiny & \tiny {Inf} & \tiny {Inf} & \tiny {0} & \tiny {Inf} &  \tiny {Inf} & \tiny {Inf} & \tiny {Inf}& \tiny {Inf}& \tiny {Inf}  \\
	{\tiny 16202}      \tiny & \tiny {Inf} & \tiny {Inf} & \tiny {Inf} & \tiny {0} &  \tiny {Inf} & \tiny {Inf} & \tiny {Inf}& \tiny {Inf}& \tiny {Inf}  \\
	{\tiny 16115}      \tiny & \tiny {Inf} & \tiny {Inf} & \tiny {Inf} & \tiny {Inf} &  \tiny {0} & \tiny {Inf} & \tiny {Inf}& \tiny {Inf}& \tiny {Inf}  \\
	{\tiny 23983}      \tiny & \tiny {2} & \tiny {3} & \tiny {3} & \tiny {4} &  \tiny {2} & \tiny {0} & \tiny {2}& \tiny {2}& \tiny {3}  \\
	{\tiny 23985}      \tiny & \tiny {Inf} & \tiny {Inf} & \tiny {Inf} & \tiny {Inf} &  \tiny {Inf} & \tiny {Inf} & \tiny {0}& \tiny {Inf}& \tiny {Inf}  \\
	{\tiny 20217}      \tiny & \tiny {Inf} & \tiny {Inf} & \tiny {Inf} & \tiny {Inf} &  \tiny {Inf} & \tiny {Inf} & \tiny {Inf}& \tiny {0}& \tiny {Inf}  \\
	{\tiny 14052}      \tiny & \tiny {Inf} & \tiny {Inf} & \tiny {Inf} & \tiny {Inf} &  \tiny {Inf} & \tiny {Inf} & \tiny {0}& \tiny {Inf}& \tiny {0}  \\
\end{tabular}
\begin{footnotesize}
		\begin{minipage}{\linewidth}%
			\centering \small Table \ref{table:SPLengthDNetgraph} shows the lengths of shortest paths from criminal to criminal in the directed Netgraph.
		\end{minipage}%
	\end{footnotesize}
\end{table}

\begin{table}[H]
\centering
	\caption{Shortest path length from criminal to criminal in the directed BCC Netgraph}
	\label{table:SPLengthDBCCNetgraph}
\begin{tabular}{l*{12}{l}l}
              \multicolumn{14}{c} {\footnotesize Directed BCC Netgraph} \\
	 {\tiny }  		& {\tiny 686} & {\tiny687} & {\tiny 11010} & {\tiny 11009} & {\tiny 10068}  & {\tiny 12708}  &\tiny{1369}&\tiny{9994}&\tiny{8716}& {\tiny 15224} & {\tiny 15225} & {\tiny 861} & {\tiny 205} \\
	\hline
	{\tiny 686}		& {\tiny 0} & {\tiny 4} & {\tiny 3} & {\tiny 3} & {\tiny 2} & {\tiny 3} &\tiny{2} &\tiny{3}&\tiny{3}& {\tiny 2}& {\tiny 4} & {\tiny 4}& {\tiny Inf}  \\
	{\tiny 687}		& \tiny {Inf}& \tiny {0} & \tiny {Inf} & \tiny {Inf} &  \tiny {Inf} & \tiny {Inf} &\tiny{Inf}& \tiny {Inf}& \tiny {Inf}& \tiny {Inf} & \tiny {Inf} & \tiny {Inf}& \tiny {Inf} \\
	{\tiny 11010}	& \tiny {3}& \tiny {1} & \tiny {0} & \tiny {3} &  \tiny {3} & \tiny {3} &\tiny{3}&\tiny{3}&\tiny{3}& \tiny {4} & \tiny {5} & \tiny {5}& \tiny {Inf} \\
	{\tiny 11009}	& \tiny {Inf}& \tiny {Inf} & \tiny {Inf} & \tiny {0} &  \tiny {Inf} & \tiny {Inf} &\tiny{Inf}& \tiny {Inf}& \tiny {Inf}& \tiny {Inf} & \tiny {Inf} & \tiny {Inf}& \tiny {Inf} \\
	{\tiny 10068}	& \tiny {2}& \tiny {4} & \tiny {3} & \tiny {3} &  \tiny {0} & \tiny {3} &\tiny{3}&\tiny{2}&\tiny{2}& \tiny {3} & \tiny {5} & \tiny {5}& \tiny {Inf} \\
	{\tiny 12708}	& \tiny {Inf}& \tiny {Inf} & \tiny {Inf} & \tiny {Inf} &  \tiny {Inf} & \tiny {0} &\tiny{Inf}&\tiny{Inf}&\tiny{Inf}& \tiny {Inf} & \tiny {Inf} & \tiny {Inf}& \tiny {Inf} \\
	{\tiny 1369}		& \tiny {1}& \tiny {4} & \tiny {3} & \tiny {3} &  \tiny {2} & \tiny {1} &\tiny{0}&\tiny{3}&\tiny{3}& \tiny {3} & \tiny {3} & \tiny {2}& \tiny {Inf} \\
	{\tiny 9994}		& \tiny {Inf}& \tiny {Inf} & \tiny {Inf} & \tiny {Inf} &  \tiny {Inf} & \tiny {Inf} &\tiny{Inf}&\tiny{0}&\tiny{Inf}& \tiny {Inf} & \tiny {Inf} & \tiny {Inf}& \tiny {Inf} \\
	{\tiny 8716}		& \tiny {Inf}& \tiny {Inf} & \tiny {Inf} & \tiny {Inf} &  \tiny {Inf} & \tiny {Inf} &\tiny{Inf}&\tiny{Inf}&\tiny{0}& \tiny {Inf} & \tiny {Inf} & \tiny {Inf}& \tiny {Inf} \\
	{\tiny 15224}          & \tiny {3}& \tiny {4} & \tiny {3} & \tiny {3} &  \tiny {3} & \tiny {3} &\tiny{3}&\tiny{3}&\tiny{3}& \tiny {0} & \tiny {4} & \tiny {4}& \tiny {Inf} \\
	{\tiny 15225}          & \tiny {Inf}& \tiny {Inf} & \tiny {Inf} & \tiny {Inf} &  \tiny {Inf} & \tiny {Inf} &\tiny{Inf}&\tiny{Inf}&\tiny{Inf}& \tiny {Inf} & \tiny {0} & \tiny {Inf}& \tiny {Inf} \\
	{\tiny 861}              & \tiny {Inf}& \tiny {Inf} & \tiny {Inf} & \tiny {Inf} &  \tiny {Inf} & \tiny {Inf} &\tiny{Inf}&\tiny{Inf}&\tiny{Inf} & \tiny {Inf} & \tiny {Inf} & \tiny {0}& \tiny {Inf} \\
	{\tiny 205}              & \tiny {Inf}& \tiny {Inf} & \tiny {Inf} & \tiny {Inf} &  \tiny {Inf} & \tiny {Inf} &\tiny{Inf}&\tiny{Inf}&\tiny{Inf}& \tiny {Inf} & \tiny {Inf} & \tiny {Inf}& \tiny {0} \\
\end{tabular}
\begin{footnotesize}
		\begin{minipage}{\linewidth}
                \captionsetup{singlelinecheck=off,justification=centerfirst}
			\centering  \small Table \ref{table:SPLengthDBCCNetgraph} shows the lengths of shortest paths from criminal to criminal in the directed BCC Netgraph.
		\end{minipage}%
	\end{footnotesize}
\end{table}

From Tables \ref{table:SPLengthDNetgraph} and \ref{table:SPLengthDBCCNetgraph} it is clear that, in the directed graphs under consideration, only a few criminals have directed paths connecting them to other criminals. If we make the broad assumption that an email sent from A to B implies an \emph{undirected} relationship between A and B then the graphs become undirected. In this case, in BCC Netgraph, 12 of the 13 accounts associated with criminals belong to the same connected component and a path can be found from one criminal's account to another's (see Table \ref{table:SPundirectedLengthBCCNetgraph}). The exception is the account associated with A. Khan (adnankkhan@hotmail.com) which belongs to a separate component. The assumption regarding reciprocal relationship seems most appropriate for the trust network (BCC Netgraph) where if A includes B as a BCC recipient there is a personal trust relationship implied between A and B that we assume is reciprocated to some degree. 

The shortest path lengths between the criminals in the undirected graphs are shown in Tables \ref{table:SPundirectedLengthNetgraph} and \ref{table:SPundirectedLengthBCCNetgraph}. The average path length values of the undirected Netgraph and the undirected BCC Netgraph are 3.264676 and 5.033507 respectively. The average path length of the criminals in the undirected Netgraph and the undirected BCC Netgraph are 2.93 and 3.65 respectively; lower than the average path length of the entire graphs. 
 
\begin{table}[H]
\centering
	\caption{Shortest path length from criminal to criminal in the undirected Netgraph}
	\label{table:SPundirectedLengthNetgraph}
\begin{tabular}{l*{8}{l}l}
              \multicolumn{10}{c} {\footnotesize Undirected Netgraph } \\
	 {\tiny }  & {\tiny 1472} & {\tiny17589} & {\tiny 17588} & {\tiny 16202} & {\tiny 16115}  & {\tiny 23983} & {\tiny 23985} & {\tiny 20217} & {\tiny 14052} \\
	\hline
	{\tiny 1472}          & {\tiny 0} & {\tiny 3} & {\tiny 2} & {\tiny 3} & {\tiny 2} & {\tiny 2} & {\tiny 3 }& {\tiny 2 }& {\tiny 2}  \\
	{\tiny 17589}       & \tiny {3 }& \tiny {0} & \tiny {2} & \tiny {4} &  \tiny {3} & \tiny {3} & \tiny {3} & \tiny {3} & \tiny {3}  \\
          {\tiny 17588}      \tiny & \tiny {2} & \tiny {2} & \tiny {0} & \tiny {4} &  \tiny {3} & \tiny {3} & \tiny {4} & \tiny {3} & \tiny {3}  \\
	{\tiny 16202}      \tiny & \tiny {3} & \tiny {4} & \tiny {4} & \tiny {0} &  \tiny {3} & \tiny {3} & \tiny {4}& \tiny {4}& \tiny {4}  \\
	{\tiny 16115}      \tiny & \tiny {2} & \tiny {3} & \tiny {3} & \tiny {3} &  \tiny {0} & \tiny {2} & \tiny {3}& \tiny {2}& \tiny {2}  \\
	{\tiny 23983}      \tiny & \tiny {2} & \tiny {3} & \tiny {3} & \tiny {3} &  \tiny {2} & \tiny {0} & \tiny {3}& \tiny {2}& \tiny {2}  \\
	{\tiny 23985}      \tiny & \tiny {3} & \tiny {3} & \tiny {4} & \tiny {4} &  \tiny {3} & \tiny {3} & \tiny {0}& \tiny {4}& \tiny {4}  \\
	{\tiny 20217}      \tiny & \tiny {2} & \tiny {3} & \tiny {3} & \tiny {4} &  \tiny {2} & \tiny {2} & \tiny {4}& \tiny {0}& \tiny {3}  \\
	{\tiny 14052}      \tiny & \tiny {2} & \tiny {3} & \tiny {3} & \tiny {4} &  \tiny {2} & \tiny {2} & \tiny {4}& \tiny {3}& \tiny {0}  \\
\end{tabular}
\begin{footnotesize}
		\begin{minipage}{\linewidth}%
			\small \centering Table \ref{table:SPundirectedLengthNetgraph} above shows the lengths of shortest paths from criminal to criminal in undirected Netgraph.
		\end{minipage}%
	\end{footnotesize}
\end{table}

\begin{table}[H]
\centering
	\caption{Shortest path length from criminal to criminal in the undirected BCC Netgraph}
	\label{table:SPundirectedLengthBCCNetgraph}
\begin{tabular}{l*{12}{l}l}
              \multicolumn{14}{c} {\footnotesize  Undirected BCC Netgraph} \\
	 {\tiny }  		& {\tiny 686} & {\tiny687} & {\tiny 11010} & {\tiny 11009} & {\tiny 10068}  & {\tiny 12708} & {\tiny 1369} & {\tiny 9994} & {\tiny 8716} & {\tiny 15224} & {\tiny 15225} & {\tiny 861} & {\tiny 205} \\
	\hline
	{\tiny 686}		& {\tiny 0} & {\tiny 4} & {\tiny 3} & {\tiny 3} & {\tiny 2} & {\tiny 2} & {\tiny 1 }& {\tiny 2 }& {\tiny 3} & {\tiny    3}& {\tiny 5} & {\tiny 4}& {\tiny Inf}  \\
	{\tiny 687}		& \tiny {4}& \tiny {0} & \tiny {1} & \tiny {3} &  \tiny {4} & \tiny {5} & \tiny {4} & \tiny {5} & \tiny {5}  & \tiny {5} & \tiny {6} & \tiny {7}& \tiny {Inf} \\
          {\tiny 11010}		& \tiny {3}& \tiny {1} & \tiny {0} & \tiny {2} &  \tiny {3} & \tiny {4} & \tiny {3} & \tiny {4} & \tiny {4}  & \tiny {4} & \tiny {5} & \tiny {6}& \tiny {Inf} \\
	{\tiny 11009}	& \tiny {3}& \tiny {3} & \tiny {2} & \tiny {0} &  \tiny {3} & \tiny {4} & \tiny {3} & \tiny {4} & \tiny {4}  & \tiny {3} & \tiny {5} & \tiny {6}& \tiny {Inf} \\
	{\tiny 10068}	& \tiny {2}& \tiny {4} & \tiny {3} & \tiny {3} &  \tiny {0} & \tiny {2} & \tiny {2} & \tiny {2} & \tiny {2}  & \tiny {2} & \tiny {5} & \tiny {5}& \tiny {Inf} \\
	{\tiny 12708}	& \tiny {2}& \tiny {5} & \tiny {4} & \tiny {4} &  \tiny {2} & \tiny {0} & \tiny {1} & \tiny {3} & \tiny {3}  & \tiny {3} & \tiny {5} & \tiny {4}& \tiny {Inf} \\
	{\tiny 1369}		& \tiny {1}& \tiny {4} & \tiny {3} & \tiny {3} &  \tiny {2} & \tiny {1} & \tiny {0} & \tiny {2} & \tiny {3}  & \tiny {3} & \tiny {4} & \tiny {3}& \tiny {Inf} \\	
	{\tiny 9994}		& \tiny {2}& \tiny {5} & \tiny {4} & \tiny {4} &  \tiny {2} & \tiny {3} & \tiny {2} & \tiny {0} & \tiny {2}  & \tiny {3} & \tiny {5} & \tiny {5}& \tiny {Inf} \\
	{\tiny 8716}		& \tiny {3}& \tiny {5} & \tiny {4} & \tiny {4} &  \tiny {2} & \tiny {3} & \tiny {3} & \tiny {2} & \tiny {0}  & \tiny {2} & \tiny {5} & \tiny {6}& \tiny {Inf} \\
	{\tiny 15224}         & \tiny {3}& \tiny {5} & \tiny {4} & \tiny {3} &  \tiny {2} & \tiny {3} & \tiny {3} & \tiny {3} & \tiny {2}  & \tiny {0} & \tiny {4} & \tiny {5}& \tiny {Inf} \\
	{\tiny 15225}         & \tiny {5}& \tiny {6} & \tiny {5} & \tiny {5} &  \tiny {5} & \tiny {5} & \tiny {4} & \tiny {5} & \tiny {5}  & \tiny {4} & \tiny {0} & \tiny {6}& \tiny {Inf} \\
	{\tiny 861}         	& \tiny {4}& \tiny {7} & \tiny {6} & \tiny {6} &  \tiny {5} & \tiny {4} & \tiny {3} & \tiny {5} & \tiny {6}  & \tiny {5} & \tiny {6} & \tiny {0}& \tiny {Inf} \\
	{\tiny 205}         	& \tiny {Inf}& \tiny {Inf} & \tiny {Inf} & \tiny {Inf} &  \tiny {Inf} & \tiny {Inf} & \tiny {Inf} & \tiny {Inf} & \tiny {Inf}  & \tiny {Inf} & \tiny {Inf} & \tiny {Inf}& \tiny {0} \\
\end{tabular}
\begin{footnotesize}
		\begin{minipage}{\linewidth}%
			\small \centering Table \ref{table:SPundirectedLengthBCCNetgraph} above shows the lengths of shortest paths from criminal to criminal in undirected BCC Netgraph.
		\end{minipage}%
	\end{footnotesize}
\end{table}

The distance between any two criminals in the undirected Netgraph ranges from 2-4 (see Table \ref{table:SPundirectedLengthNetgraph}) while in the undirected BCC Netgraph it ranges from 1-7 (see Table \ref{table:SPundirectedLengthBCCNetgraph}). Using the shortest paths' lengths count, there are variations in the connections formed between criminals in the undirected BCC Netgraph compared to the undirected Netgraph. In the undirected BCC Netgraph, there are some direct links between certain criminals, for example Andrew Fastow (BCCNet. ID 686) to Lea Fastow (11010), Andrew Fastow (686) to Ben Glisan (1369) and from Michael Kopper (12708) to Ben Glisan (1369). The emphasis of this paper is on the BCC Netgraph. In the next subsection, we describe our shortest paths network search algorithm briefly.

\subsection{Criminal network formation methods}
\label{DescriptionofSPNSA}

Past research shows that a criminal community can be formed using certain pre-defined rules. For example in \citep{al2012}, a set of people belong to the same community if their names appear together in a document while \citep{anwar214} group people into a community based on overlapping interests across different chat sessions. Our shortest paths network search algorithm is used to form a relationship network between suspects as a basis for an investigation \citep{MagalingamP2014}.  Unlike \citet{al2012} and \citet{anwar214}, our algorithm does not restrict community membership to those with similarity in email content. By retaining duplicate email addresses (unlike \citet{al2012}), we show that these could indicate secret trusted connections. Our network link doesn't depend on overlapping interest such as in \citet{anwar214} but depends on a node's associations with particular central nodes and its links to known suspects. In our algorithm, the node or edge with highest centrality value is used. Girvan and Newman \citep{girvan02, newman04} remove the edge with highest betweenness score till they find different sub-networks or communities. In more recent work, \citep{ferrara2014} use a log analysis tool that adapts several community detection algorithms as well as utilising modular optimisation as done in Girvan and Newman algorithm \citep{girvan02} and Newman's fast algorithm \citep{newman2004fast} to detect criminal organisations using phone call networks. Unlike the log analysis tool of \citep{ferrara2014} that implements Girvan and Newman's algorithm, we retain the central nodes, using them to form sub-networks of interest.

The shortest paths network search algorithm (SPNSA) is described in detail in \citep{MagalingamP2014}. Firstly the algorithm requires a `feed', a number of nodes of interest, whether these are known to be suspected criminals or otherwise regarded as relevant or important. For the Enron email network, each email account is used as a feed. For example, if a criminal has two email accounts, both the accounts are used in the feed list to represent that one criminal (see Table \ref{table:MLCriminals}). Then the algorithm works by isolating a particular ego network, and within that ego network, identifying two central nodes, one with highest betweenness centrality and the other with highest eigenvector centrality. These nodes are named the Middle Man (MM) and the Most Influential (MI) respectively. Within the  same ego network, the algorithm proceeds to extract the shortest paths from the ego to the central nodes, as well as from the ego to other nodes of interest in the list and finally, from the other nodes of interest to the central nodes. The three steps are repeated for every node of interest by selecting each in turn as the ego. The results of the various extractions are combined to form a new sub-network. In Magalingam (2014), we studied the use of SPNSA on the directed BCC email network containing emails with 1 and 2 recipients bcc-ed. We found that the sub-network formed using SPNSA suggested possible people to investigate between known criminals and financial managers. This new sub-network could be used by an investigator as a preliminary investigative network \citep{MagalingamP2014}. In this paper, we apply the SPNSA to analyse larger subsets of the Enron dataset than those studied in \citep{MagalingamP2014} in addition to comparing its performance against known community detection algorithms. Before applying the SPNSA, we use the different community detection algorithms to discover the various criminals' communities in the Netgraph and BCC Netgraph.

\section{Discovering criminals' community using community detection algorithms}
\label{Discovery of Criminals}

Many authors have used algorithms to analyse community structure and to consequently identify groups or sub-networks. An example of this is the use of network modularity in community detection algorithms \citep{girvan02, newman04}. Communities exist when a graph consists of sets of nodes in tightly knit groups joined together by weaker connections between these groups \citep{girvan02, newman2010}. The link structure and node attributes are the common components used in community detection algorithms. Girvan and Newman \citep{girvan02, newman04} repeatedly calculate edge betweenness, each time removing the edge with highest betweenness score such that as the graph becomes disconnected, the components represent each of communities. Other link based community detection approaches can be found in \citep{Radicchi20-04}.

A different way of identifying communities is by using the node based approach called agglomerative algorithms \citep{blondel2008}. Pons and Latapy \citep{Pons20-06} introduce a random walk concept which picks nodes from a network based on a fixed distance between two nodes. EAGLE is a software algorithm created by Shen et al. \citep{Shen20-09} that follows certain steps to form a community. First, it adapts the maximal clique calculation introduced by Bron and Kerbosch \citep{Bron:1973}, then removes the subordinate maximal clique.  The algorithm then calculates the similarity between each pair of nodes in the clique, merges them into a new community, finds the similarity of the new community by comparing with an already existing community, repeating the steps until only one community remains.

Fastgreedy community detection \citep{Clauset20-04} uses a modularity optimization algorithm by first computing the fraction of within-community edges in a network, then subtracting from it the expected fraction of edges in a randomized version of the same network with same degree distribution. A nonzero value above 0.3 is considered a good measurement for the density of links inside communities \citep{Clauset20-04} \citep{blondel2008}. Walktrap community detection merges similar nodes that are obtained using short random walks into a group \citep{Pons20-06}. The leading eigenvector community detection algorithm implements the modularity optimization algorithm. It computes the modularity matrix and the eigenvector of the matrix. It then divides the community based on the positive or negative sign of the elements in the eigenvector. If the large elements have the same sign, then the network has no community structure \citep{Nman20-06}.

The results produced by applying these different methods to the Enron dataset to identify criminals' communities are discussed next. The community detection methods used are $k$-Neighbourhood, Fastgreedy, Walktrap and Leading Eigenvector algorithms all of which are available in the R igraph tool. Due to the connections between criminals being more visible in the undirected graph compared to the directed graph (See Tables \ref{table:SPLengthDNetgraph}, \ref{table:SPLengthDBCCNetgraph}, \ref{table:SPundirectedLengthNetgraph} and \ref{table:SPundirectedLengthBCCNetgraph}) and since the majority of the community algorithms can only be applied to undirected graphs, we use the undirected Netgraph and BCC Netgraph for this exercise. We will later compare (in Section \ref{Result of SPN on undirected Netgraph and BCC Netgraph}) these results with those obtained by using our shortest paths network search algorithm.

\subsection{$k$-Neighbourhood detection}
\label{Criminalneighbourhoods}

We first compute the total degree of each criminal in both the undirected Netgraph and BCC Netgraph. Table \ref{table:MLCriminalsdegree} shows the values. The neighbourhood function has been used previously to form a subgraph and identify the nearest link from a criminal node \citep{savage2014, yasin2014}. Using Table \ref{table:MLCriminalsdegree}, we find all the neighbours of Andrew Fastow (686), (the first criminal in our list) in the undirected BCC Netgraph at a distance of 1 to 4 and form the networks.

	
	\begin{table}[H]
		\centering
		\protect\caption{{\small Enron money laundering criminals' degree }}
		\label{table:MLCriminalsdegree}
		
		\vspace{0.2cm}
		
		\centering{}%
	
		\begin{tabular}{|l|c|c|c|c|}
	
		\hline
		\small Name & \small Net. ID & \small Degree & \small BCCNet. ID & \small Degree \tabularnewline
		\hline
		\small Andrew Fastow & \small 1472  & \small 261 & \small 686 & \small 25\tabularnewline
		\hline
		\small Andrew Fastow & \small - & \small - & \small 687 & \small 1 \tabularnewline
		\hline
		\small Lea Fastow & \small 17589 & \small 3 & \small 11010 & \small 4 \tabularnewline
		\hline
		\small Lea Fastow & \small 17588 & \small 4 & \small 11009 & \small 4\tabularnewline
		\hline
		\small Kevin Hannon & \small 16202 & \small 1 & \small 10068 & \small 36\tabularnewline
		\hline
		\small Kenneth Rice & \small 16115 & \small 11 & \small 9994 & \small 4 \tabularnewline
		\hline
		\small Rex Shelby & \small 23983 & \small 97 & \small 15224 & \small 21\tabularnewline
		\hline
		\small Rex Shelby & \small 23985 & \small 2 & \small 15225 & \small 2\tabularnewline
		\hline
		\small A. Khan & \small - & \small - & \small 205 & \small 2\tabularnewline
		\hline
		\small Michael Kopper & \small 20217 & \small 40 & \small 12708 & \small 6\tabularnewline
		\hline
		\small Ben Glisan & \small - & \small - & \small 1369 & \small 105\tabularnewline
		\hline
		\small Joe Hirko & \small 14052 & \small 11 & \small 8716 & \small 2\tabularnewline
		\hline
		\small S. Yaeger & \small - & \small - & \small 861 & \small 1\tabularnewline
		\hline	
	\end{tabular}
	
	\vspace{1 em}
	\begin{footnotesize}
		\begin{minipage}{\linewidth}%
			\small \centering Table \ref{table:MLCriminalsdegree} shows the total degree of each criminal in Netgraph and BCC Netgraph.
		\end{minipage}%
	\end{footnotesize}
\end{table}

\begin{figure}[H]
    \centering
    \subfigure[1-N network (26 nodes, 149 edges) ]
    {
        \includegraphics[height=3.0in, width=3.0in]{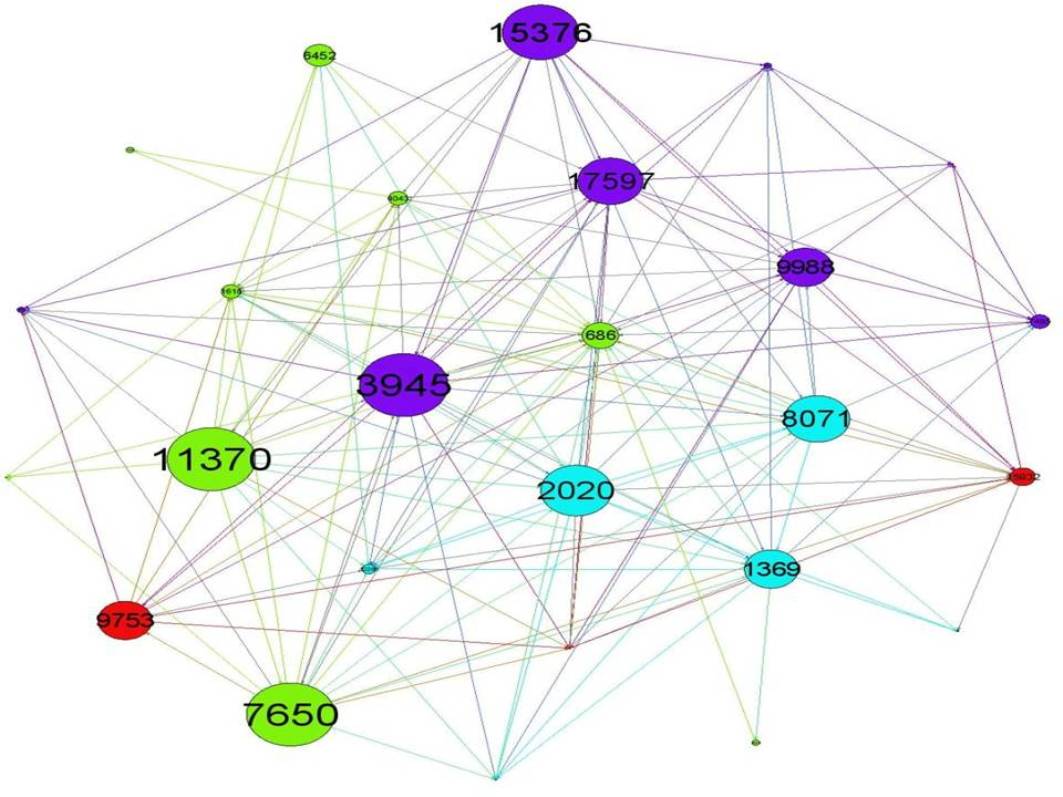}
        \label{fig:third_sub}
    }
    \hspace{1em}
    \subfigure[2-N network (1,559 nodes, 15,853 edges)]
    {
        \includegraphics[height=3.0in, width=3.0in]{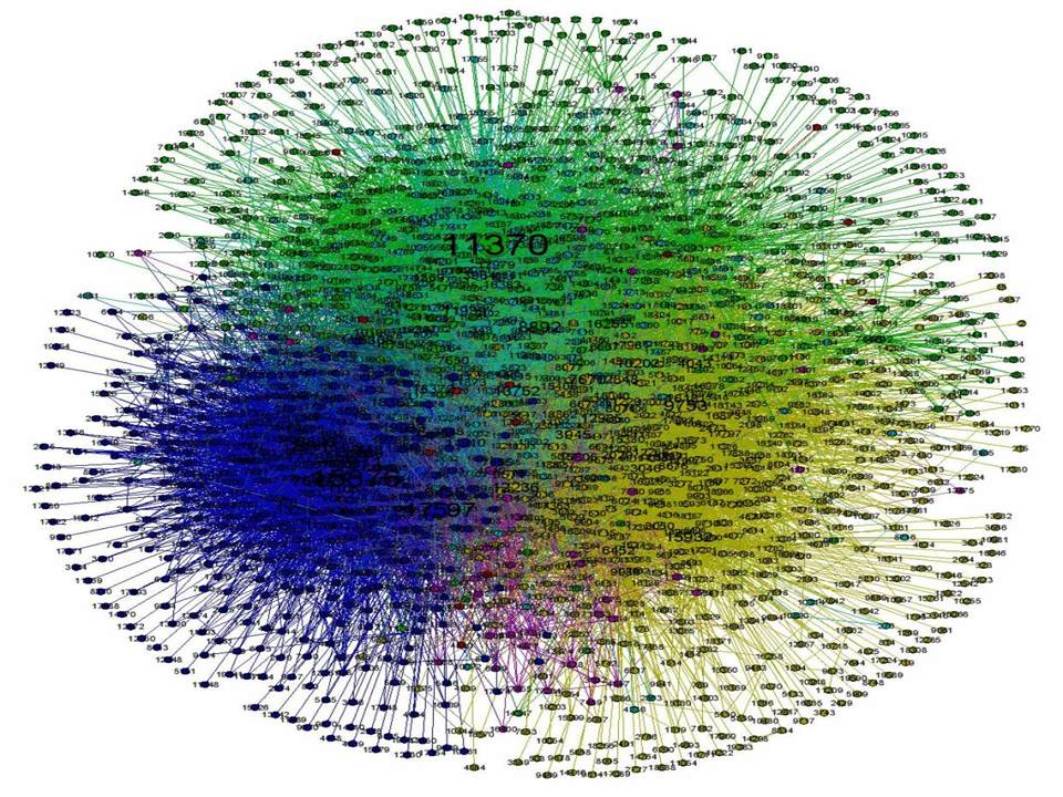}
        \label{fig:fourth_sub}
    }
    \\
    \subfigure[3-N network (8,433 nodes, 49,086 edges)]
    {
       \includegraphics[height=3.0in, width=3.0in]{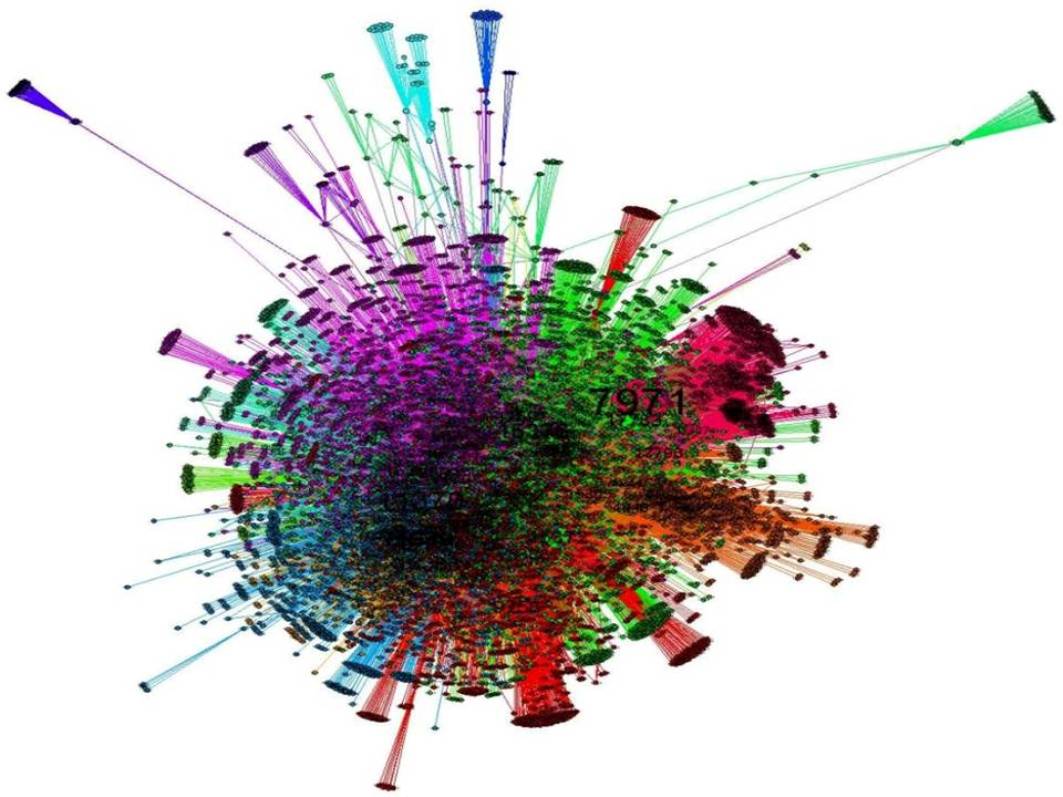}
        \label{fig:fifth_sub}
    }
    \hspace{1em}
    \subfigure[4-N network (14,916 nodes, 60,593 edges)]
    {
       \includegraphics[height=3.0in, width=3.0in]{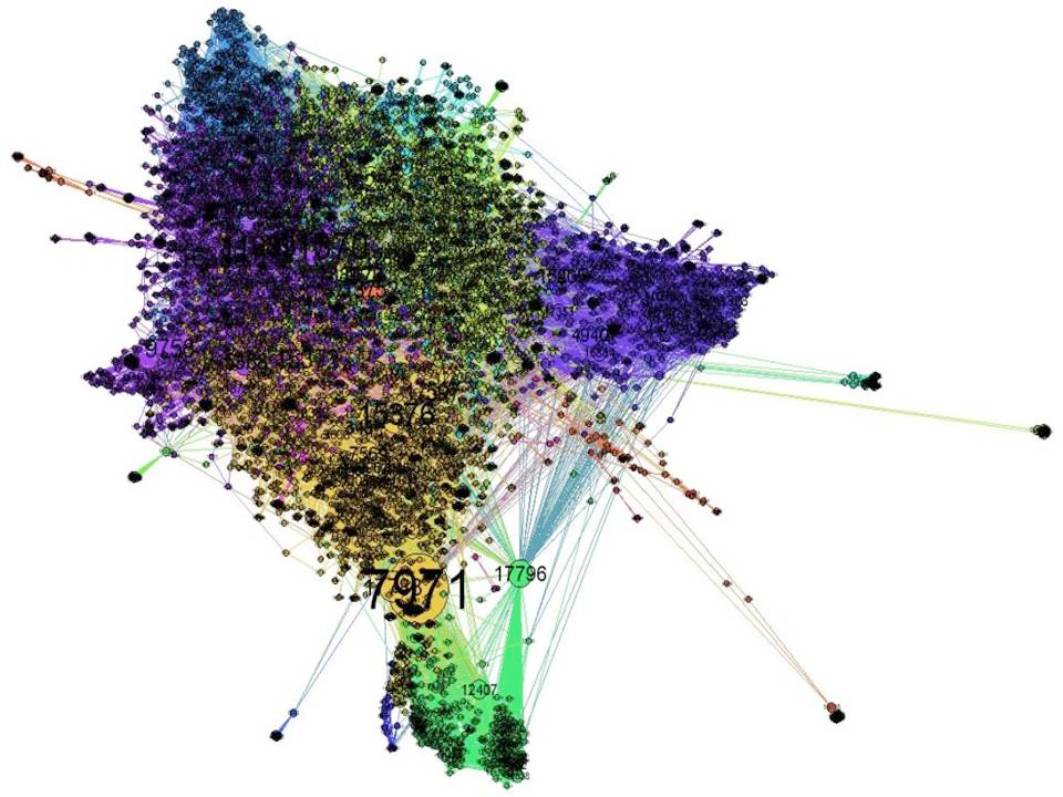}
        \label{fig:sixth_sub}
    }
\captionsetup{singlelinecheck=off,justification=justified}
    \caption{\small The figures above show the networks formed by using 1,2,3,4-neighbourhood (1,2,3,4-N) of Andrew Fastow (686) in the undirected BCC Netgraph respectively. In the 1-neighbourhood of Andrew Fastow only one other criminal was found, Ben Glisan (1369). The 2, 3, and 4- neighbourhood networks are clearly too dense to be able to identify other criminals easily.}
    \label{fig:NAFBCCConnection}
\end{figure}

Figure \ref{fig:NAFBCCConnection} shows the networks found using 1,2,3,4-neighbourhood of Andrew Fastow (686) in the undirected BCC Netgraph respectively. 

In the 1-neighbourhood network of Andrew Fastow only one criminal was found, Ben Glisan (1369). According to Tables \ref{table:SPLengthDNetgraph},\ref{table:SPLengthDBCCNetgraph}, \ref{table:SPundirectedLengthNetgraph},\ref{table:SPundirectedLengthBCCNetgraph} using either of the Andrew Fastow's email account (BCCNet. ID 686 or 687) and either of the undirected or directed BCC Netgraphs, the one or two neighbourhoods would only rarely contain the other known criminals. The size of the network becomes bigger as the neighbourhood increases. The same method when applied to the undirected Netgraph, also produces large graphs that are difficult to explore. Clearly, using these dense network graphs, it is difficult to analyse a criminal's occurrence and connections with other nodes.

\subsection{Community detection algorithms in R $igraph$}
\label{RCommunityDetections}

In the next two sub-sections, \ref{Results of TO and CC Email network} and \ref{Results of BCC Email network}, we compare the number of criminals, communities and connection of criminals to other nodes found using the community detection algorithms: Fastgreedy \citep{Clauset20-04}, Walktrap \citep{Pons20-06} and Leading Eigenvector \citep{Nman20-06}. As all three of these algorithms require undirected graphs, we use the undirected Netgraph and BCC Netgraph for this experiment.

\subsubsection{Results of undirected Netgraph}
\label{Results of TO and CC Email network}

Here we look at the results of applying the community detection algorithms to the undirected Netgraph. Applying the Fastgreedy community detection algorithm gives 15 communities. These communities range in size from 5,803 nodes to just 2 nodes. 6 out of 10 criminals were found in the second largest community that had 5,163 nodes and 22,936 edges.  One criminal, Kevin Hannon (16202) appeared in a much smaller group consisting of 230 nodes and 781 edges.

Next, the Walktrap community detection algorithm was applied to the undirected Netgraph with the length of the random walk being 10 steps. 530 communities were found by the algorithm, with the first community detected consisting of 3,126 nodes and 12,462 edges, and again contained 6 of the 10 criminals (See Table \ref{table:Community IDs}). It was the second largest community formed by Walktrap.  The largest community had 7,935 nodes while smallest one had just 1 node.

The third community detection algorithm used was the Leading Eigenvector. This detection algorithm detected 2 communities with the largest one containing 26,025 nodes and the smallest 2 nodes. All 7 criminals appeared in the largest community but the criminals were found to be isolated (See Table \ref{table:Community IDs}).

\subsubsection{Results of undirected BCC Netgraph}
\label{Results of BCC Email network}

The community detection algorithms were next applied to the undirected BCC network graph. The BCC Netgraph contains 65,532 edges and 19,716 nodes. The Fastgreedy algorithm found 832 communities, finding a number of small communities with less than 6 nodes each. 5 criminals were found to be in the largest community that had 2,195 nodes and 7,903 edges; Andrew Fastow (BCCNet. ID 686), Lea Fastow (BCCNet. ID 11010, BCCNet. ID 11009), Kevin Hannon (BCCNet. ID 10068), Ben Glisan (BCCNet. ID 1369) and Kenneth Rice (BCCNet. ID 9994). Ben Glisan (BCCNet. ID 1369) had the highest degree in this community. Meanwhile, Rex Shelby (15224, 15225) and S. Yaeger (861) belonged to the second largest community with 2,142 nodes and 8,222 edges. The criminal who was in one of the smaller communities was Joe Hirko (8716).

Using the Walktrap community detection algorithm produced 1,773 communities from the undirected BCC Netgraph. The largest community had 1,493 nodes and the smallest one had just 1 node. Seven out of 10 criminals happened to exist in the same community that had 1,254 nodes (See Table \ref{table:Community IDs}). The Leading Eigenvector community detection algorithm found 719 communities in the BCC Netgraph. It ranged from the largest community with 15,792 nodes and the smallest with 1 node. Most of the criminals appeared in the largest community. In the Table \ref{table:Community IDs}, we list the communities where the criminals belong to and the community size that we identified manually.

	\begin{table}[H]
		\centering
		\protect\caption{{\small Criminals Found in Different Communities}}
		\label{table:Community IDs}
		
		\vspace{0.2cm}
		
		\centering{}%
		
		\begin{tabular}{|p{1.8cm}|p{1.5cm}|p{1.5cm}|p{1.5cm}||p{1.5cm}|p{1.5cm}|p{1.5cm}|p{1.5cm}|}

		\hline
		 \tiny Net. ID & \tiny FG Com. ID & \tiny WT Com. ID & \tiny LEC Com. ID & \tiny BCCNet. ID & \tiny FG Com. ID & \tiny WT Com. ID & \tiny LEC Com. ID
              \tabularnewline
		\hline
		\tiny 1472 & \tiny \{7, 5163\} &\tiny\{1, 3126\} &\tiny \{1, 26025\} &\tiny 686 &\tiny \{5, 2195\} &\tiny \{36, 1254\} &\tiny \{1, 2001\} \tabularnewline
            \hline
		-& -  & - & - &\tiny 687 & \tiny \{5, 2195\} &\tiny \textcolor{red}{\{594, 2\}} &\tiny \{719, 15792\} \tabularnewline
		\hline
		\tiny 17589 &\tiny \{7, 5163\} &\tiny \{1, 3126\} & \tiny \{1, 26025\} & \tiny 11010 & \tiny \{5, 2195\} &\tiny \textcolor{red}{\{594, 2\}} &\tiny  \{719, 15792\}\tabularnewline
		\hline
		\tiny 17588 &\tiny \{7, 5163\} & \tiny \{1, 3126\} &\tiny \{1, 26025\} &\tiny11009 &\tiny \{5, 2195\} &\tiny \{36, 1254\}&\tiny \{719, 15792\} \tabularnewline
		\hline
		\tiny16202 &\tiny \{10, 230\} &\tiny \{29, 228\} &\tiny \{1, 26025\} &\tiny 10068 & \tiny \{5, 2195\} &\tiny \{36, 1254\} &\tiny \{719, 15792\}\tabularnewline		
            \hline
		\tiny16115 & \tiny \{7, 5163\} &\tiny \{1, 3126\} &\tiny \{1, 26025\} &\tiny 9994 & \tiny \{5, 2195\} & \tiny \{36, 1254\} & \tiny \{719, 15792\} \tabularnewline
		\hline
		\tiny 23983 & \tiny \{7, 5163\} & \tiny \{1, 3126\} & \tiny \{1, 26025\} &\tiny 15224 & \tiny \{3, 2142\} & \tiny \{36, 1254\} & \tiny \{719, 15792\} \tabularnewline
		\hline	
		\tiny 23985 & \tiny \{7, 5163\} & \tiny \{4, 7935\} & \tiny \{1, 26025\} & \tiny 15225 & \tiny \{3, 2142\} &\tiny \textcolor{red}{\{1365, 1\}} &\tiny\{719, 15792\} \tabularnewline
		\hline
		- & - & - & - & \tiny 205 & \tiny \textcolor{red}{\{224, 3\}} & \tiny \textcolor{red}{\{1030, 3\}} & \tiny \{719, 15792\}\tabularnewline
		\hline
	   \tiny 20217 &\tiny \{7, 5163\} &\tiny \{1, 3126\} & \tiny\{1, 26025\} & \tiny12708 &\tiny \{2, 2113\} &\tiny \{36, 1254\} & \tiny\{1, 2001\}\tabularnewline
		\hline
	    - & - & - & - & \tiny 1369 & \tiny \{5, 2195\} & \tiny \{45, 1493\} & \tiny \{719, 15792\}\tabularnewline
		\hline
		 \tiny 14052 & \tiny \{7, 5163\} & \tiny \{1, 3126\} & \tiny \{1, 26025\} & \tiny 8716 & \tiny \{17, 561\} & \tiny \{36, 1254\} & \tiny \{719, 15792\}\tabularnewline
		 \hline
		- & - & - & - & \tiny 861 & \tiny \{3, 2142\} & \tiny \{54, 1101\} & \tiny \{719, 15792\}\tabularnewline
		\hline
		 \tiny Total community & \tiny 15 & \tiny 530 & \tiny 2 & - & \tiny 832 & \tiny 1773 & \tiny 719 \tabularnewline
		 \hline
		\end{tabular}
	\vspace{1 em}
	\begin{footnotesize}
		\begin{minipage}{\linewidth}%
\captionsetup{singlelinecheck=off,justification=justified}
			\small Table \ref{table:Community IDs} shows community IDs to which each criminal belongs. The community and the size of each community is represented in curly brackets as \{$i_{th}$ community, size\}. The title of each column are Net. ID (Netgraph ID), FG Com. ID (Fastgreedy Community ID), WT Com. ID (Walktrap Community ID) and LEC Com. ID (Leading Eigenvector Community ID). The total number of communities formed is shown at the bottom of the table. The communities with smallest number of nodes, 1-3 nodes are highlighted in red.
		\end{minipage}%
	\end{footnotesize}
\end{table}

\subsubsection{Discussion of results obtained by R $igraph$ community detection algorithms}
\label{Discussion}

The network partitioning using leading eigenvector to form communities seems not to be very effective for either the undirected Netgraph or BCC Netgraph, as the biggest community contains almost all the nodes. Some abnormal network structures were found in the communities of certain criminals (see Table \ref{table:Community IDs}). From the results of the Walktrap algorithm (highlighted in red), Andrew Fastow (687) and Lea Fastow (11010) belong to a small group of just two nodes, forming a community on their own. A. Khan (205) also belongs to a community of just three nodes. A. Khan is linked to two other nodes with email addresses; toriarules@aol.com and mmorales@arnel.com. These emails addresses are found to be external emails that do not belong to the Enron company email group. The Fastgreedy algorithm results in Rex Shelby (15225) being isolated in a community of his own, numbered 1365. All of these abnormalities occurred in the undirected BCC Netgraph. We also found more communities appearing in the undirected BCC Netgraph compared to the undirected Netgraph. The total number of communities formed using each detection algorithm is also shown in the last row of Table \ref{table:Community IDs}.

\begin{figure}[H]
    \centering
    \subfigure[Community numbered 5 using Fast Greedy algorithm]
    {
        \includegraphics[height=2.0in, width=2.8in]{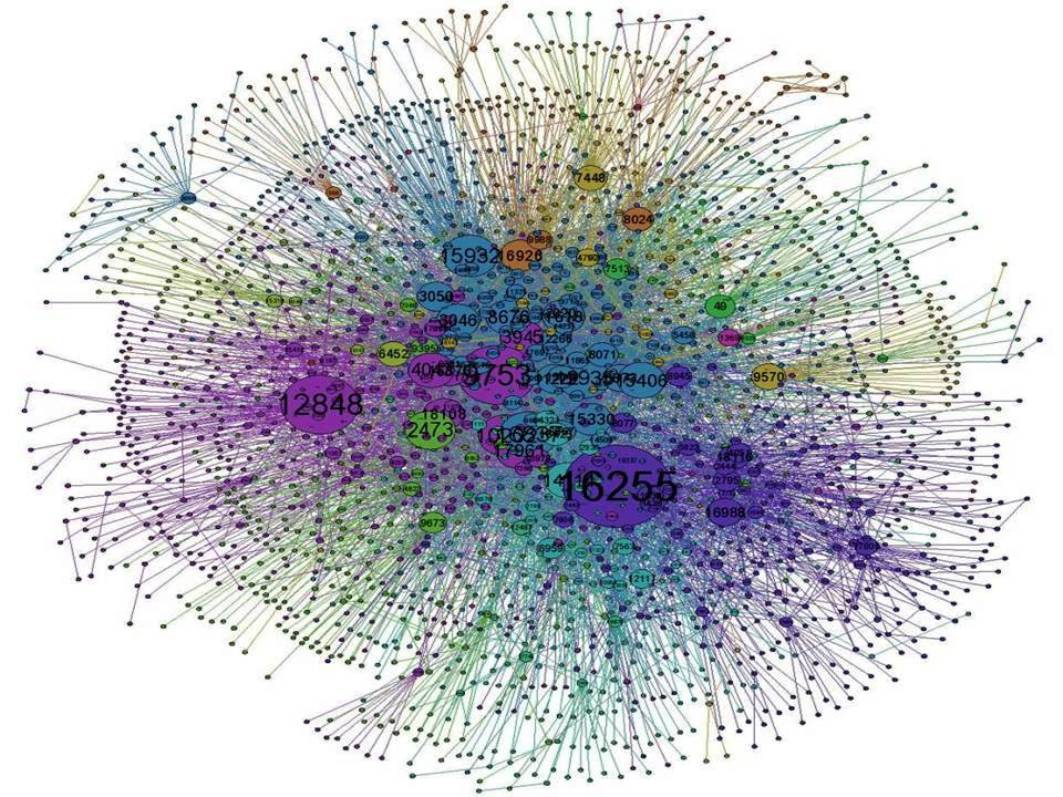}
        \label{fig:seventh_sub}
    }
    \subfigure[Community numbered 36 using WalkTrap algorithm ]
    {
        \includegraphics[height=2.0in, width=2.8in]{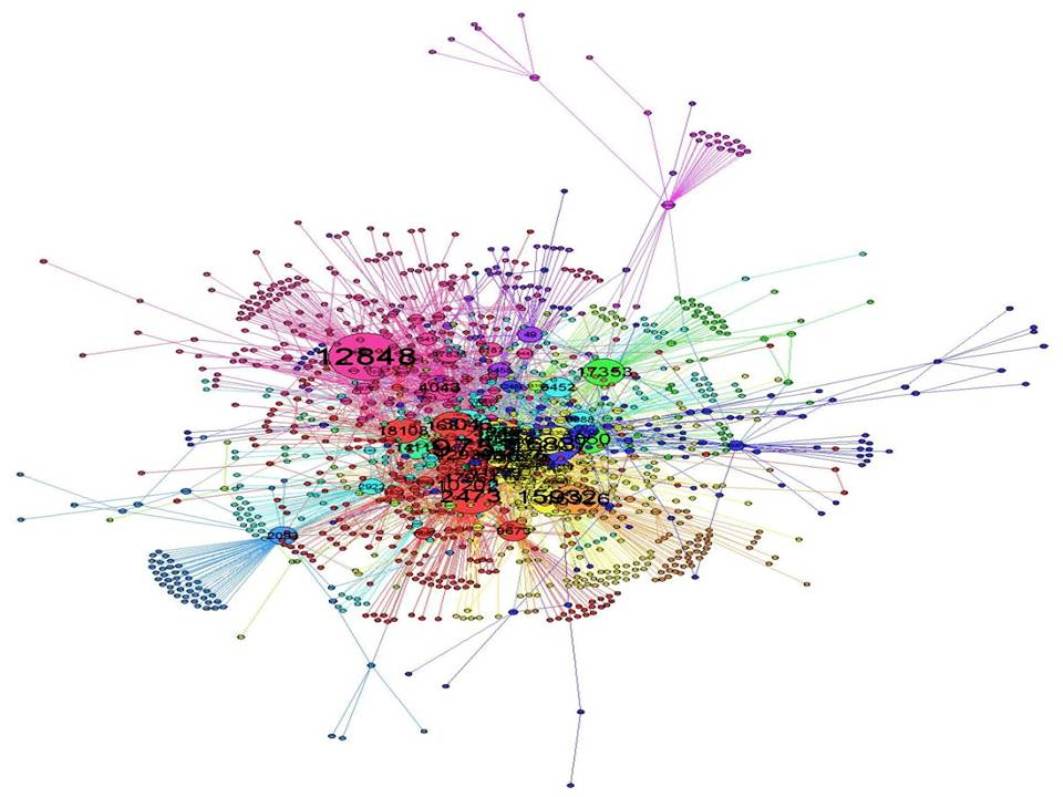}
        \label{fig:eighth_sub}
    }
\captionsetup{singlelinecheck=off,justification=justified}
    \caption{\small Figures above shows the community numbered 5 (2195 nodes) using fast greedy algorithm and community numbered 36 (1254 nodes) using walktrap algorithm on undirected BCC Netgraph.}
    \label{fig:CommFromDetectionAlgorithm}
\end{figure}

The community that contains the most criminals; community numbered 5 using Fastgreedy algorithm and community numbered 36 using Walktrap algorithm on the undirected BCC Netgraph were extracted. The detection using Walktrap algorithm on the BCC Netgraph yields the best result, some criminals appear in a small community on their own and 7 of 1254 members were criminals, but the result shows enormous number of nodes and links; networks (as shown in Figure \ref{fig:CommFromDetectionAlgorithm}) an investigator would need to analyse in order to find any connections between criminals and other nodes.

Another method called clique percolation community detection developed by \citep{palla2005} was also used to identify the Enron criminal community. We found that the clustering coefficient values for both the undirected Netgraph and BCC Netgraph were so low that the nodes did not converge to form communities. A high average clustering coefficient to a respective random network is needed to form sub-networks or clusters \citep{palla2005, hills2009}. In the next section, we apply our shortest paths network search algorithm to all four networks, the directed and undirected, Netgraph and BCC Netgraph.

\section{Application of Shortest Path Network Search Algorithm}
\label{SPNSA}

In \citep{MagalingamP2014}, the shortest paths network search algorithm (SPNSA) was used to identify a trust network from a network of emails that have 1 and 2 bcc-ed recipients respectively to start an investigation. Here, we apply our SPNSA to the directed and undirected Netgraph and the BCC Netgraph. Different from \citep{MagalingamP2014}, the directed and undirected BCC Netgraph used here contain all bcc-ed recipients. In this section, all the criminals in Table \ref{table:MLCriminalsdegree} are used as the feed for SPNSA. There are 7 criminals in the Netgraph and 10 criminals in BCC Netgraph (see Table \ref{table:MLCriminalsdegree}).

\subsection{Application of SPNSA on directed Netgraph and BCC Netgraph}
\label{Result of SPN on directed Netgraph and BCC Netgraph}

\begin{figure}[H]
	\DeclareGraphicsExtensions{.pdf,.png,.jpg}
	\centering
	\includegraphics[width=4.0in]{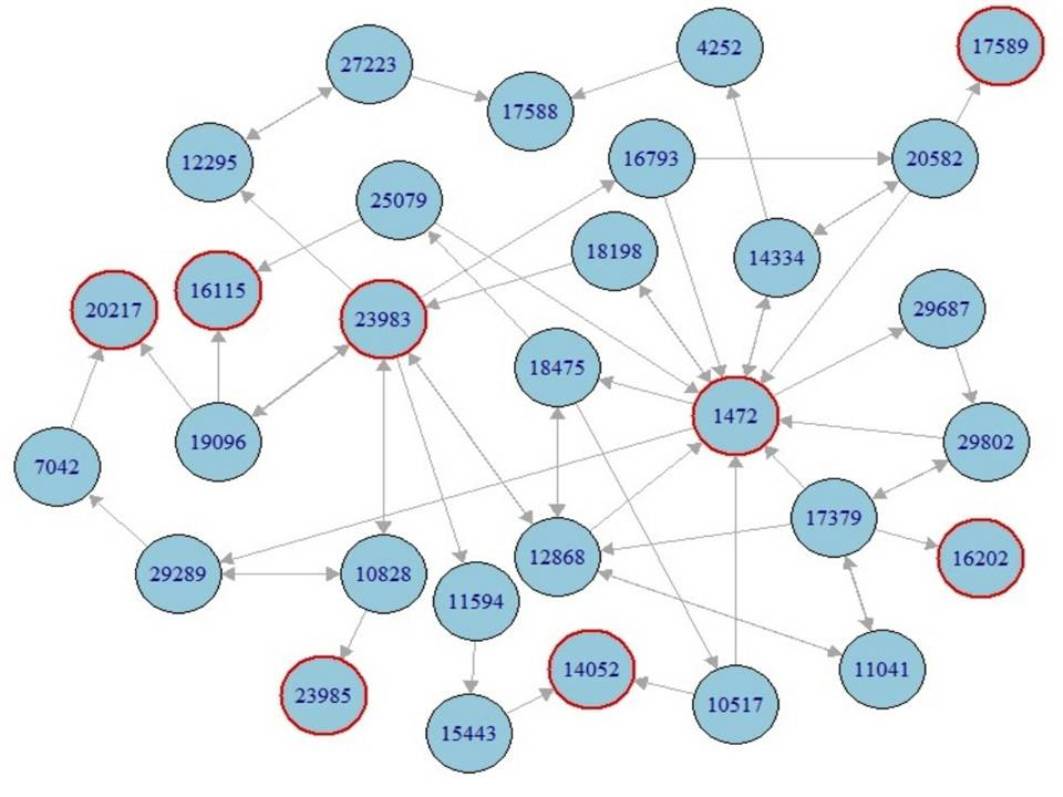}
\captionsetup{singlelinecheck=off,justification=justified}	
\caption{The shortest paths network formed using the directed Netgraph. All 7 criminals were found. The network formed is sparse and criminals' connections can be easily identified. This sub-network contains 30 nodes. The nodes highlighted in red represent the criminals with Net. ID in Table \ref{table:MLCriminals}.}
	\label{fig:SPNConnection}
\end{figure}

Applying the SPNSA to the directed Netgraph results in all 7 criminals occurring in the extracted shortest paths network except for one email ID of Lea Fastow (Net. ID 17589) (see Figure \ref{fig:SPNConnection}). Lea Fastow's Net. ID 17589 that represents her second email address didn't appear in the sub-network because the node doesn't have an out-component that builds paths to other criminals or the central nodes.

\begin{figure}[H]
	\DeclareGraphicsExtensions{.pdf,.png,.jpg}
	\centering
	\includegraphics[width=4.5in]{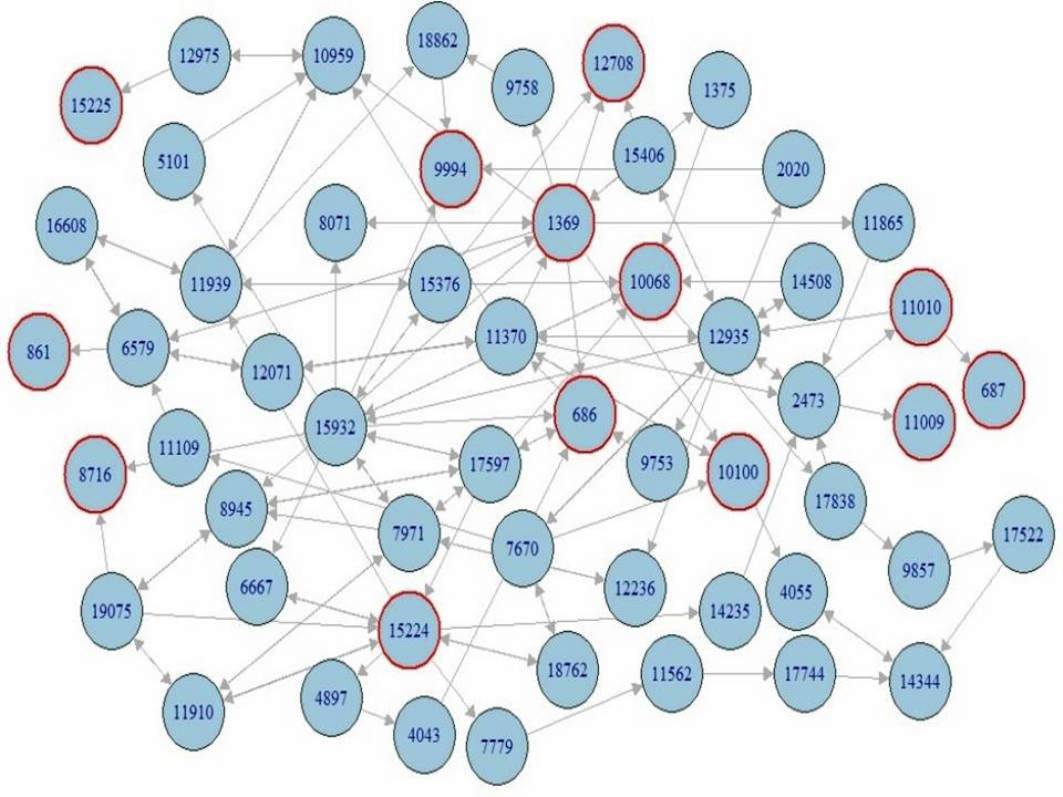}
\captionsetup{singlelinecheck=off,justification=justified}	
\caption{The shortest paths network formed using directed BCC Netgraph. 2 out of 10 criminals were lost. The network formed is sparse and criminals' connections can be identified. The number of nodes in this sub-network is 55. The nodes highlighted in red represent the criminals with BCCNet. ID in Table \ref{table:MLCriminals}.}
	\label{fig:SPNBCCConnection}
\end{figure}

The sub-network of the directed BCC Netgraph captured using SPNSA is shown in Figure \ref{fig:SPNBCCConnection}. This network contains 8 out of the 10 criminals. The two other criminals did not have any connection to other criminals or to the MM or MI, thus did not occur in this sub-network.

\subsection{Application of SPNSA on the undirected Netgraph and the undirected BCC Netgraph}
\label{Result of SPN on undirected Netgraph and BCC Netgraph}

\begin{figure}[H]
	\DeclareGraphicsExtensions{.pdf,.png,.jpg}
	\centering
	\includegraphics[width=4.0in]{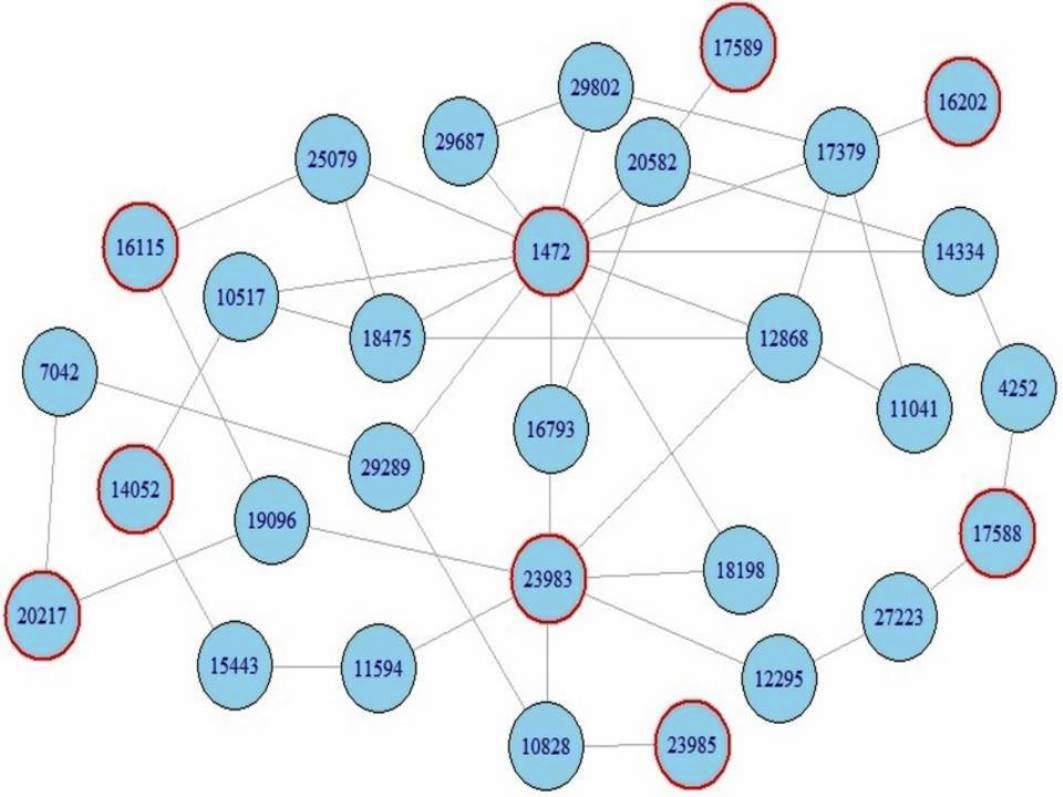}
\captionsetup{singlelinecheck=off,justification=justified}	
\caption{The shortest paths network formed using undirected Netgraph. No criminal IDs were lost. The network formed is sparse and criminals' connections can be identified. The size of this sub-network is 30 nodes. The nodes highlighted in red represent the criminals with Net. ID in Table \ref{table:MLCriminals}.}
	\label{fig:SPNonundirectedNetgraph}
\end{figure}

The SPNSA was then applied to the undirected Netgraph and the result is depicted in Figure \ref{fig:SPNonundirectedNetgraph}. Again, all seven criminals occurred in this graph. This time all double email Net. IDs are captured in this sub-network.

\begin{figure}[H]
	\DeclareGraphicsExtensions{.pdf,.png,.jpg}
	\centering
	\includegraphics[width=4.5in, height=3.0in]{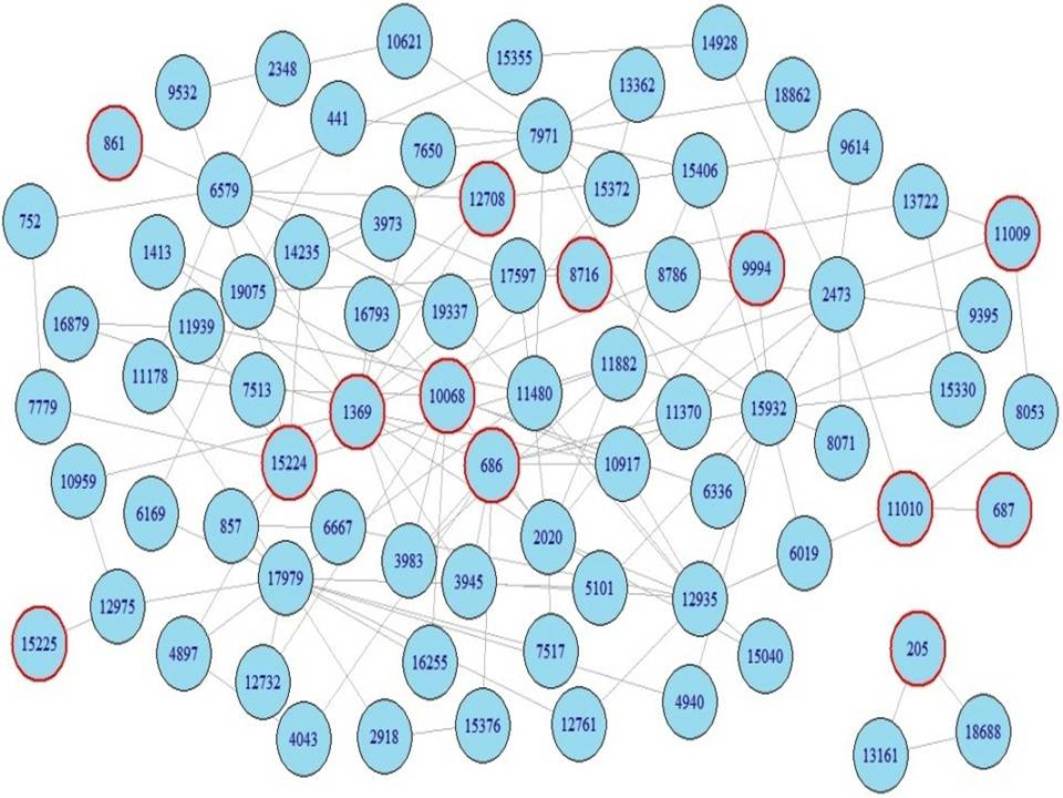}
\captionsetup{singlelinecheck=off,justification=justified}	
\caption{The shortest paths network formed using undirected BCC Netgraph. No criminal were lost. The network formed is sparse and criminals' connections can be identified. The size of this sub-network is 74 nodes and another small component of 3 nodes. The nodes highlighted in red represent the criminals with BCCNet. ID in Table \ref{table:MLCriminals}.}
	\label{fig:SPNonundirectedBCCNetgraph}
\end{figure}

We also applied the SPNSA to the undirected BCC Netgraph and the result is shown in Figure \ref{fig:SPNonundirectedBCCNetgraph}. The result shows that, in this case, SPNSA is able to identify two connected components, with all 10 criminals. When compared to the result obtained from the undirected Netgraph, application of SPNSA to the undirected BCC Netgraph gives better results as it is able to show all the criminals' connections and has more connected components (see Figure \ref{fig:SPNonundirectedBCCNetgraph}).

The criminals and their links can be clearly seen in the results obtained (see Figures \ref{fig:SPNConnection}, \ref{fig:SPNBCCConnection},  \ref{fig:SPNonundirectedNetgraph} and \ref{fig:SPNonundirectedBCCNetgraph}). The undirected BCC Netgraph yields the most number of criminals in the shortest paths network and connected components. A comparison of the results of using R $igraph$ community detection algorithms with the result of shortest paths network search algorithm (SPNSA) applied to the undirected BCC Netgraph (see Figure \ref{fig:SPNonundirectedBCCNetgraph}) is documented in Table \ref{table:ComparisonigraphAlgoandSPNSA}.

\begin{table}[H]
\centering
	\caption{Comparison between results found using R igraph community detection algorithms and SPNSA}
	\label{table:ComparisonigraphAlgoandSPNSA}
\begin{tabular}{|p{1.6cm}|p{3.6cm}|p{3.9cm}|p{2.5cm}|p{2.5cm}|}
 \cline{1-5}
  & \multicolumn{2}{c|} {\tiny Community Detection Algorithm} & \multicolumn{2}{c|} {\tiny  Shortest Paths Network Search Algorithm}\\
  \cline{2-5}
  &\tiny  Undirected Netgraph    &\tiny Undirected BCC Netgraph    &\tiny Undirected Netgraph    &\tiny Undirected BCC Netgraph\\
  \hline
  \tiny Community distribution & \tiny Nature: large and difficult to explore. Community Size: 250 < nodes < 26,100. & \tiny Nature: large and difficult to explore. Community Size: 1,000 < nodes < 16,000.  & \tiny Nature: sparse to investigate and explore. Community Size: 30 nodes. & \tiny Nature: sparse to investigate and explore. Two components occurred, one of size 74 nodes and the other has 3 nodes.\\
  \hline
  \tiny Abnormalities & \tiny Andrew Fastow's external email add. (687) doesn't exist. & \tiny Walktrap community detection: found Andrew Fastow's external email add. (687) and Lea Fastow (11010) appeared in the same small community of size 2 nodes.& \tiny Andrew Fastow's external email add. (687) doesn't exist. & \tiny Found a direct connection between Andrew Fastow (687) and Lea Fastow (11010) that emerged in a community of size 74 nodes.\\
  \cline{2-5}
   & \tiny A. Khan (205) doesn't exist &\tiny Fastgreedy and Walktrap community detection: found A. Khan (205) belongs to a small community of size 3. & \tiny A. Khan (205) doesn't exist. & \tiny Found A. Khan  (205) belong to a small isolated community of size 3. \\
   \hline
  \tiny Total criminals & \tiny Detection Algorithm: Fastgreedy - 6/10 criminals in \{7, 5163\}. Walktrap - 6/10 criminals in \{1, 3126\}. & \tiny Detection Algorithm: Fastgreedy: 5/10 criminals in \{5, 2195\}. Walktrap: 7/10 criminals in \{36, 1254\}. & \tiny 8/10 criminals. & \tiny All 10 criminals.\\
  \hline
\end{tabular}

\begin{footnotesize}
	\vspace{1 em}
	\begin{minipage}{\linewidth}%
\captionsetup{singlelinecheck=off,justification=justified}	
		\footnotesize Table \ref{table:ComparisonigraphAlgoandSPNSA} shows the comparison between results found using R igraph community detection algorithms and SPNSA. The community formed by SPNSA is small and suitable for investigation. The 3-clique connection is formed as a separate network component and the nodes that are connected to the criminal can be easily identified. In the row giving total criminals, (\{7, 5163\}) refers to \{$i_{th}$ community, size\} and the same follows for others.
	\end{minipage}%
\end{footnotesize}
\end{table}

\section{Crime investigation methods using SPNSA}
\label{CrimeInvestigationMethod}

\citet{anwar2012} analysed their algorithm's performance by implementing three different scenarios based on the availability of information. Similar to \citep{anwar2012}, here we specify certain ways an investigator could extract criminal subgraphs using the shortest paths network search algorithm for a preliminary investigation. In this section, the undirected BCC Netgraph is chosen instead of the undirected Netgraph due to three reasons found through our experiments in section \ref{IdentifyingNetworkMeasure} and the comparison between results in Table \ref{table:ComparisonigraphAlgoandSPNSA}; more connections were detected between the criminals in the undirected BCC Netgraph (see Tables \ref{table:SPLengthDNetgraph}, \ref{table:SPLengthDBCCNetgraph}, \ref{table:SPundirectedLengthNetgraph} and \ref{table:SPundirectedLengthBCCNetgraph}), cliques of two or three criminals were found in the undirected BCC Netgraph (See Table \ref{table:ComparisonigraphAlgoandSPNSA}) and the most number of criminals were found in the undirected BCC Netgraph (See Table \ref{table:ComparisonigraphAlgoandSPNSA}).

\subsection{Extracting sub-networks using Non-Criminals}
\label{Extractingsubgraphs}

\begin{figure}[H]
	\DeclareGraphicsExtensions{.pdf,.png,.jpg}
	\centering
	\includegraphics[width=5.6in, height=2.8in]{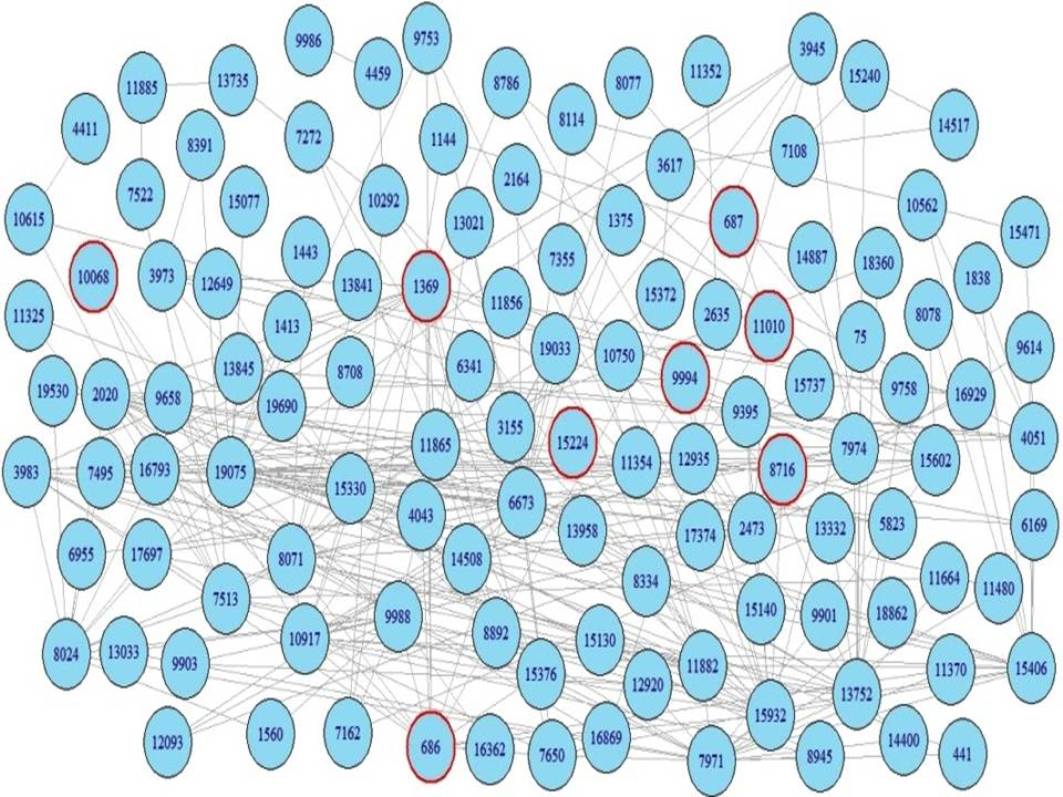}
\captionsetup{singlelinecheck=off,justification=justified}
	\caption{Enron undirected BCC Netgraph shortest paths network using all top managers as algorithm feed. The nodes highlighted in red are all criminals.}
	\label{fig:ShortestPathsNetworkUsingAllManagers}
\end{figure}

At an intial stage of a criminal investigation, an investigator may or may not have all or any of the criminals' details. The investigator could start the investigation with a suitable group of people. The investigator will be able to feed in as many as necessary of the Enron managers' or suspects' node IDs in to the algorithm to form a network for their investigation. We simulate such a scenario by feeding in to the algorithm all the top managers in the Enron to obtain their communication network from the undirected BCC Netgraph. The result is shown in Figure \ref{fig:ShortestPathsNetworkUsingAllManagers}.

When compared to the network graph obtained in Figure \ref{fig:SPNonundirectedBCCNetgraph}, 7 out of 10 criminals are found by the algorithm this time around. Apart from the managers known to be criminals (see Section \ref{Money laundering criminals}), the SPNSA also extracts 4 other criminals; Lea Fastow (11010), Kevin Hannon (10068), Rex Shelby (15224) and Joe Hirko (8716) with this top manager feed test.

\begin{figure}[H]
	\DeclareGraphicsExtensions{.pdf,.png,.jpg}
	\centering
	\includegraphics[width=4.0in]{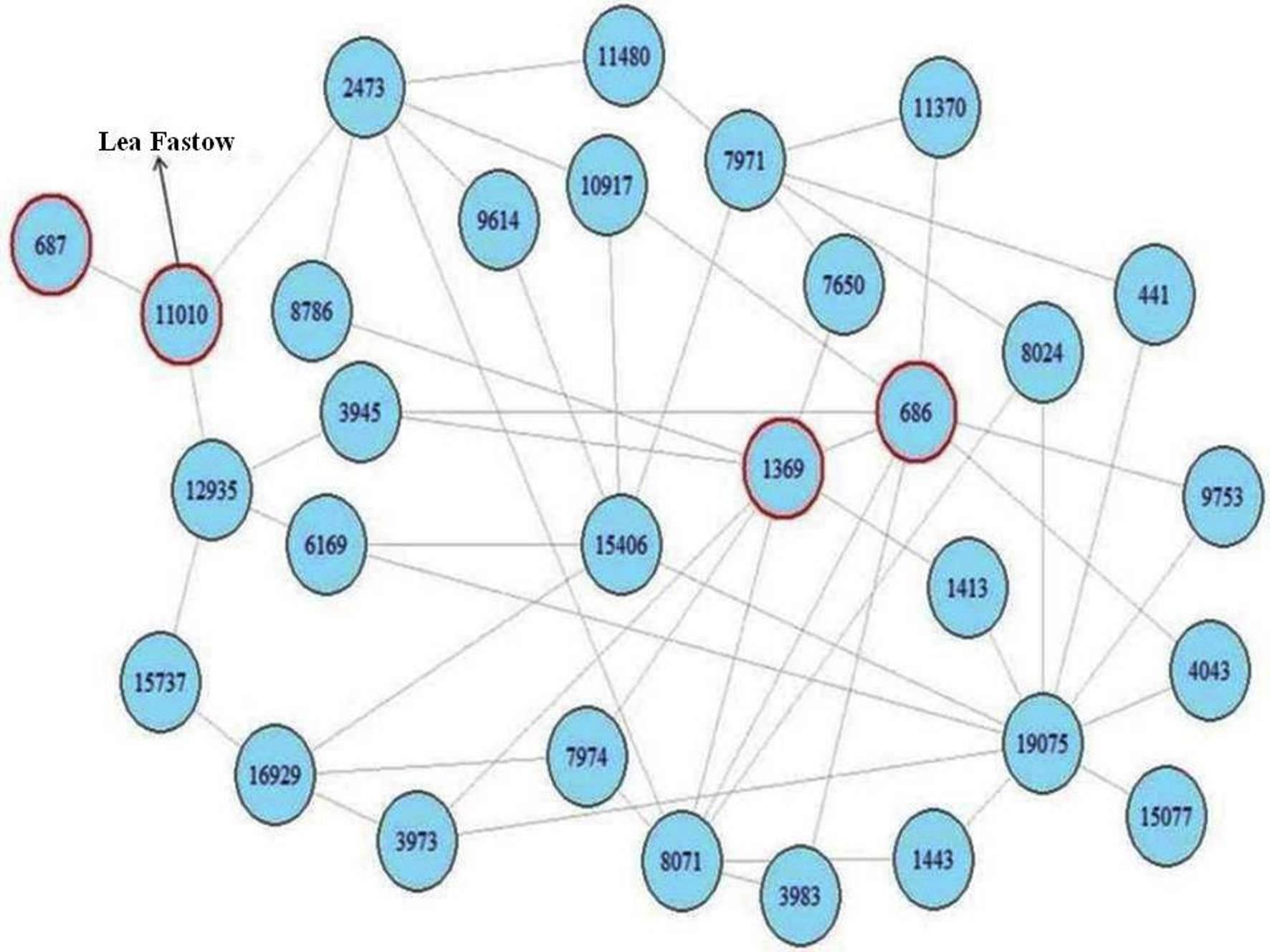}
\captionsetup{singlelinecheck=off,justification=justified}
	\caption{Enron undirected BCC Netgraph shortest paths network using all financial managers as algorithm feed. The nodes (686, 687 and 1369) highlighted in red are financial managers who are also criminals. The other criminal who was found is Lea Fastow (11010).}
	\label{fig:ShortestPathsNetworkUsingFinancialManagers}
\end{figure}

Next a shortest paths network is formed using only financial managers. The financial managers' group is a subset of the top managers. The financial managers are the Head of Enron Global Finance, Sherron Watkins (16929), the Enron Chief Financial Officer, Andrew Fastow (686 , 687), the Enron Corporation Treasurer, Ben Glisan (1369),  the Chief Accounting Officer, Rick Causey (15077), the Chief Financial Officer of Enron after Andrew Fastow, Jeff McMahon (8071). The network formed is shown in Figure \ref{fig:ShortestPathsNetworkUsingFinancialManagers}.

When comparing the criminals found in the sub-network formed using the Finance managers (see Figure \ref{fig:ShortestPathsNetworkUsingFinancialManagers}), with those found in Figure \ref{fig:SPNonundirectedBCCNetgraph}, we see that Lea Fastow (11010) is the only other criminal found here, other than the finance managers who were also criminals.

\subsection{Extracting subgraphs using leave-one-out method}
\label{Extractingsubgraphs}

The leave-one-out method is widely used in various fields of research as a data sampling method for an algorithm \citep{cawley2004, kocaguneli2013} and can be used to estimate performance of a predictive model \citep{kocaguneli2013}. Past research \citep{shao1996} shows that one can set the number of data points to be removed from sample data and use it for validation. This is also called delete-p cross validation \citep{zhang1993}.

We name this method as leave-$C_i$-out. Leave-$C_i$-out refers to dropping one criminal ($C_i$) from the list of criminals and running the shortest paths network search algorithm on the remaining criminals in the undirected BCC Netgraph. This method is a test of the ability of the algorithm to produce sub-networks that contain the convicted criminal not included in the algorithm feed. We name the criminal that is left out during each iteration as $C_i$. A criminal who has two email accounts has two different BCCNet. IDs and during the leave-$C_i$-out process both their IDs are dropped from the algorithm feed.

\begin{figure}[H]
	\DeclareGraphicsExtensions{.pdf,.png,.jpg}
	\centering
	\includegraphics[width=4.0in]{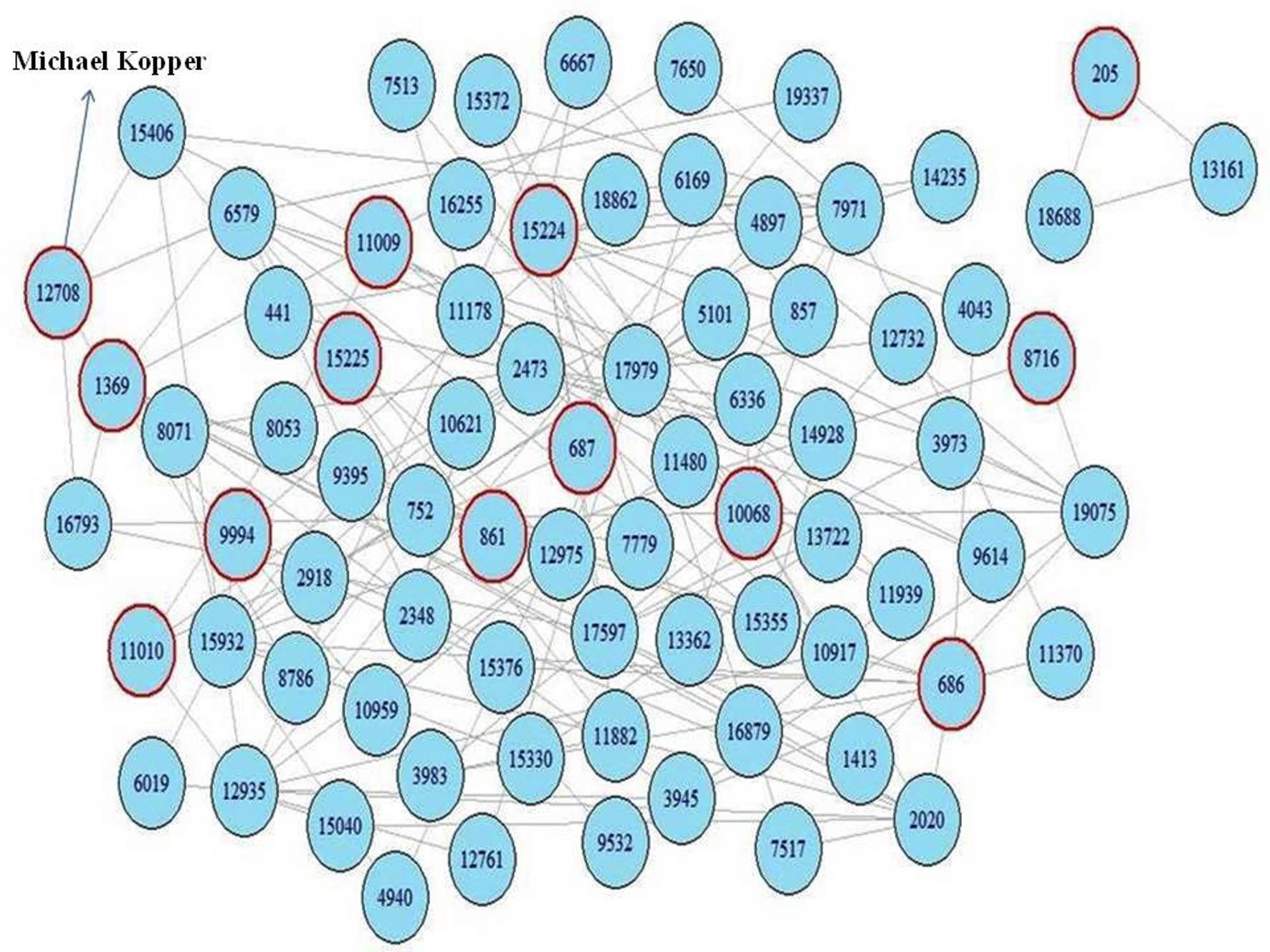}
\captionsetup{singlelinecheck=off,justification=justified}
	\caption{Enron undirected BCC Netgraph shortest paths network when Michael Kopper (12708) is left out from the criminal feed list. The nodes highlighted in red are all criminals.}
	\label{fig:ShortestPathsNetworkLMKO}
\end{figure}

The results of the leave-$C_i$-out method are given in Table \ref{table:LOO}. In 5 out of 9 cases the criminal who is left out occurs in the network formed by the shortest paths algorithm. For example, we leave out Michael Kopper (12708) from the feed (list of criminals) and run the algorithm. The result obtained is a sub-network that contains Michael Kopper and the connections of Michael Kopper (see Figure \ref{fig:ShortestPathsNetworkLMKO}).

\begin{table}[H]
		\centering
		\protect\caption{{\small Leave-$C_i$-out Method}}
		\label{table:LOO}
		
		\vspace{0.2cm}
		
		\centering{}%
	
	\begin{tabular}{|l|l|l|}
	    \hline
		 \footnotesize $C_i$ Out & \footnotesize BCCNet. ID & \footnotesize $C_i$ occur \tabularnewline
	    \hline
		\footnotesize Andrew Fastow & \footnotesize 686, 687 & \mbox{\xmark} \tabularnewline
		\hline
		\footnotesize Lea Fastow & {\footnotesize 11010, 11009} & \mbox{\checkmark} \tabularnewline
		\hline
		\footnotesize  Kevin Hannon & {\footnotesize 10068} & \mbox{\xmark} \tabularnewline
		\hline
		\footnotesize  Kenneth Rice & {\footnotesize 9994} & \mbox{\xmark} \tabularnewline
		\hline
		\footnotesize Rex Shelby & {\footnotesize 15224, 15225} & \mbox{\checkmark} \tabularnewline
		\hline
		\footnotesize  Michael Kopper & \footnotesize 12708 & \mbox{\checkmark} \tabularnewline
		\hline
		\footnotesize  Ben Glisan & {\footnotesize 1369} & \mbox{\checkmark} \tabularnewline
		\hline
		\footnotesize  Joe Hirko & {\footnotesize 8716} & \mbox{\xmark} \tabularnewline
		\hline
		\footnotesize  S. Yaeger  & {\footnotesize 861} & \mbox{\checkmark} \tabularnewline
		\hline
	\end{tabular}
	\vspace{1 em}
	\begin{footnotesize}
		\begin{minipage}{\linewidth}%
\captionsetup{singlelinecheck=off,justification=justified}
			\small Table \ref{table:LOO} shows the result for Leave-$C_i$-out method. It is a test to see if the criminals that we left out still occur in the network formed by the shortest paths algorithm. $C_i$ represents each criminal. The sign `\checkmark'  indicates the $C_i$ appeared in the shortest paths network community while `\xmark'  are given to $C_i$ who do not appear in the extracted network. A. Khan (BCCNet. ID 205) was not included in this table, because, as shown in Figure \ref{fig:SPNonundirectedBCCNetgraph}, A. Khan (205) had no connections to other criminals, and further, appeared in a separate component containing 3 nodes.
		\end{minipage}%
	\end{footnotesize}
\end{table}

\section{Conclusion}
The work presented in this paper contains the efficacy of the implementation of our shortest paths network search algorithm on larger dataset and the searches. The existing community detection algorithms in igraph did show the number of communities but an investigator would need to manually check the community to which a criminal belongs. Retrieving the neighbourhood sub-networks of the criminals in the community, as identified by the existing community detection algorithms, resulted in dense networks, which were hard to visualise and possibly, even harder to analyse.  The criminals' connections in these sub-networks were hard to view.

Our shortest paths network search algorithm (SPNSA) clearly shows the criminals' connections to other nodes in all the sub-networks it extracted. Three different investigation methods were tested using the SPNSA; when the investigator knows all the criminals, when the investigator fails to detect one of the criminals and when the investigator is at the starting stage and doesn't have any information about the criminals. In all three scenarios, the sub-network formed by SPNSA were sparse and hence, suitable for an investigator to see the connections as well as conduct further investigations. The SPNSA algorithm was able to extract and show the abnormalities through the sub-networks formed; components that contains criminals' connections with other nodes and the 3-clique component can be easily detected.

The SPNSA allows the investigator to feed in the early suspects or suspicious entity into the criminal list of the algorithm, a function that is not available through other community detection algorithms. The quality of a criminal investigation can be improved when we can specify some inputs as in this algorithm; SPNSA could be a very useful preliminary investigation tool for an investigator.

\section*{References}
\bibliography{mybibfile}

\end{document}